\documentclass[submission,copyright,creativecommons]{eptcs}
\providecommand{\event}{GCM 2020 Post-proceedings} 

\usepackage{mdframed}
\usepackage{amsmath, amssymb}
\usepackage[scr=boondoxo]{mathalfa}
\usepackage{array, adjustbox}

\usepackage{hyperref}
\usepackage{subcaption}
\usepackage{soul}
\usepackage{tikz}
\usetikzlibrary{arrows,shapes}
\usetikzlibrary{positioning}
\usepackage{tabularx, longtable}
\usepackage{bussproofs}
\usepackage{rotating}
\usepackage{longtable}
\usepackage{stmaryrd}
\newtheorem{corollary}{Corollary}
\newtheorem{theorem}{Theorem}

\newtheorem{observation}{Observation}
\newtheorem{definition}{Definition}
\newtheorem{lemma}{Lemma}
\newtheorem{example}{Example}
  {\gdef\scalefactor{#1}\begin{center}\proofSkipAmount \leavevmode}%
  {\scalebox{\scalefactor}{\DisplayProof}\proofSkipAmount \end{center} }

\newcommand{\tuple}[1]{\langle#1\rangle}

\newcommand{\ttt}{\texttt}

\newcommand{\mtt}{\mathtt}
\newcommand{\mrm}{\mathsf}

\newcommand{\Z}{\mathbb{Z}}

\newcommand{\G}{\mathcal{G}}
\newcommand{\R}{\mathcal{R}}

\newcommand{\dder}{\Rightarrow}

\newcommand{\Sem}[1]{\llbracket{#1}\rrbracket}

\newcommand{\lst}{\mathscr{l}}
\newcommand{\mrk}{\mathscr{m}}
\newcommand{\blue}{\mathtt{blue}}
\newcommand{\none}{\mathtt{none}}
\newcommand{\red}{\mathtt{red}}
\newcommand{\green}{\mathtt{green}}
\newcommand{\any}{\mathtt{any}}
\newcommand{\dashed}{\mathtt{dashed}}
\newcommand{\grey}{\mathtt{grey}}

\newcommand{\Sat}{\vDash\,}
\newcommand{\Sata}{\vDash^\alpha\,}

\newcommand{\mV}{\mrm{m_V}}
\newcommand{\mE}{\mrm{m_E}}
\newcommand{\lV}{\mrm{l_V}}
\newcommand{\lE}{\mrm{l_E}}

\newcommand{\RG}{\rho_g(G)}

\newcommand{\E}[1]{\exists_{\mrm{#1}}}
\newcommand{\A}[1]{\forall_{\mrm{#1}}}

\newcommand{\SUCCESS}{\small{\text{SUCCESS}}}
\newcommand{\FAIL}{\small{\text{FAIL}}}

\newcommand{\Def}[3]%
	{\begin{definition}[#1]\label{#2}\normalfont
    #3\hfill$\square$
	\end{definition}}%
\newcommand{\Prop}[3]%
	{$~$\begin{proposition}[#1]\label{#2}\normalfont
    #3
	\end{proposition}}%
\newcommand{\Theo}[3]%
	{\begin{theorem}[#1]\label{#2}\normalfont
    #3
	\end{theorem}}%
\newcommand{\Ex}[3]%
	{\begin{example}[#1]\label{#2}\normalfont
    #3
	\end{example}}%
\newcommand{\Lemma}[2]%
	{\begin{lemma}\label{#1}\normalfont
    #2
	\end{lemma}}%
\newcommand{\Col}[2]%
	{$~$\\\begin{corollary}\label{#1}\normalfont
    #2
	\end{corollary}}%
\newcommand{\Obsv}[2]%
	{$~$\\\begin{observation}\label{#1}\normalfont
    #2
	\end{observation}}%

\newcommand{\pproof}[1]%
	{$~$\begin{proof}\normalfont #1 \qedhere
	\end{proof}}%
\newcommand{\Remark}[1]%
	{\begin{remark}\normalfont #1
	\end{remark}}%
\newcommand{\lkr}{\langle L\leftarrow K\rightarrow R,~\Gamma\rangle}

\title{Verifying Graph Programs\\with First-Order Logic}
\author{Gia S. Wulandari\thanks{Supported by the Indonesia Endowment Fund for Education (LPDP)}
\institute{University of York\\ York, United Kingdom}
\institute{Telkom University\\
Bandung, Indonesia}
\email{gsw511@york.ac.uk}
\and
Detlef Plump
\institute{University of York\\ York, United Kingdom}
\email{detlef.plump@york.ac.uk}
}
\def\titlerunning{Verifying Graph Programs with First Order Logic}
\def\authorrunning{G.S. Wulandari \& D. Plump}
\begin{document}
\maketitle

\begin{abstract}
We consider Hoare-style verification for the graph programming language GP 2. In previous work, graph properties were specified by so-called E-conditions which extend nested graph conditions. However, this type of assertions is not easy to comprehend by programmers that are used to formal specifications in standard first-order logic. In this paper, we present an approach to verify GP 2 programs with a standard first-order logic. We show how to construct a strongest liberal postcondition with respect to a rule schema and a precondition. We then extend this construction to obtain strongest liberal postconditions for arbitrary loop-free programs. Compared with previous work, this allows to reason about a vastly generalised class of graph programs. In particular, many programs with nested loops can be verified with the new calculus.
\end{abstract}

\section{Introduction}
\label{sec:introduction}
Various Hoare-style proof systems for the graph programming language GP\,2 have been developed by Poskitt and Plump, see for example \cite{PoskittP12,Poskitt13}. These calculi use so-called E-conditions as assertions which extend nested graph conditions \cite{Pennemann09} with support for expressions. However, a drawback of E-conditions and nested graph conditions is that they are not easy to understand by average programmers who are typically used to write formal specifications in first-order logic. To give a simple example, the following  E-condition expresses that every node is labelled by an integer: $\forall$(\begin{tikzpicture}[scale=0.5, transform shape, minimum size=.1cm,baseline,thick]
   \node[circle, draw, label=left:\scriptsize 1] (a) at (0, 0) {$\mtt{a}$};
\end{tikzpicture}$\,,\,\exists$(\begin{tikzpicture}[scale=0.6, transform shape, minimum size=.1cm,baseline,thick]
   \node[circle, draw, label=left
   :\scriptsize 1] (a) at (0, 0) {$\mtt{a}$};
\end{tikzpicture}$\,\mid\,\mtt{int(a)}))$ $\land\,\forall$(\begin{tikzpicture}[scale=0.5, transform shape, minimum size=.1cm,baseline,thick]
   \node[circle, draw, label=left:\scriptsize 1, fill=red!50] (a) at (0, 0) {$\mtt{a}$};
\end{tikzpicture}$\,,\,\exists$(\begin{tikzpicture}[scale=0.6, transform shape, minimum size=.1cm,baseline,thick]
   \node[circle, draw, label=left
   :\scriptsize 1, fill=red!50] (a) at (0, 0) {$\mtt{a}$};
\end{tikzpicture}$\,\mid\,\mtt{int(a)}))$ $\land\,\forall$(\begin{tikzpicture}[scale=0.5, transform shape, minimum size=.1cm,baseline,thick]
   \node[circle, draw, label=left:\scriptsize 1, fill=green!50] (a) at (0, 0) {$\mtt{a}$};
\end{tikzpicture}$\,,\,\exists$(\begin{tikzpicture}[scale=0.6, transform shape, minimum size=.1cm,baseline,thick]
   \node[circle, draw, label=left
   :\scriptsize 1, fill=green!50] (a) at (0, 0) {$\mtt{a}$};
\end{tikzpicture}$\,\mid\,\mtt{int(a)}))$ $\land\,\forall$(\begin{tikzpicture}[scale=0.5, transform shape, minimum size=.1cm,baseline,thick]
   \node[circle, draw, label=left:\scriptsize 1, fill=blue!40] (a) at (0, 0) {$\mtt{a}$};
\end{tikzpicture}$\,,\,\exists$(\begin{tikzpicture}[scale=0.6, transform shape, minimum size=.1cm,baseline,thick]
   \node[circle, draw, label=left
   :\scriptsize 1, fill=blue!40] (a) at (0, 0) {$\mtt{a}$};
\end{tikzpicture}$\,\mid\,\mtt{int(a)}))$ $\land\,\forall$(\begin{tikzpicture}[scale=0.5, transform shape, minimum size=.1cm,baseline,thick]
   \node[circle, draw, label=left:\scriptsize 1, fill=gray!50] (a) at (0, 0) {$\mtt{a}$};
\end{tikzpicture}$\,,\,\exists$(\begin{tikzpicture}[scale=0.6, transform shape, minimum size=.1cm,baseline,thick]
   \node[circle, draw, label=left
   :\scriptsize 1, fill=gray!50] (a) at (0, 0) {$\mtt{a}$};
\end{tikzpicture}$\,\mid\,\mtt{int(a)}))$. Having to write two quantifiers that refer to the \emph{same} object appears unnatural from the perspective of standard predicate logic where a single universal quantifier would suffice. In the logic we introduce in this paper, the above condition is simply written as $\mrm{\forall_Vx(int(\lV(x)))}$. Both E-conditions and first-order formulas tend to get lengthy in examples, but our concern with nested graph conditions is that they require a non-standard interpretation. We believe that programmers cannot be expected to think in terms of morphisms and commuting diagrams, but should be allowed to work with a type of logic that they are familiar with.

In this paper we use assertions which are conventional first-order formulas enriched with GP\,2 expressions. We believe that these assertions are easier to comprehend by programmers than E-conditions and also offer the prospect of reusing the large range of tools available for first-order logic.

To use our assertions in Hoare-style verification, we show how to construct a strongest liberal postcondition Slp($c,r$) for a given conditional rule schema $r$ and a precondition $c$. Based on this construction, we can define strongest liberal postconditions for arbitrary loop-free graph programs and preconditions. Moreover, for loop-free programs we give syntactic conditions on host graphs which express successful execution resp.\ the existence of a failing execution. With these results we obtain a verification calculus that can handle considerably more programs than the calculi in \cite{PoskittP12,Poskitt13}. In particular, many programs with nested loops can now be formally verified, which has been impossible so far.

Nevertheless, our proof calculus is not relatively complete because first-order logic is not powerful enough to express all necessary assertions. Therefore we present a semantic version of the calculus which turns out to be relatively complete. The space available for this paper does not allow us to present all technical details or the proofs of our results. These can be found in the long version \cite{WP20a}. 

The remainder of this paper is structured as follows. A brief review of the graph programming language GP\,2 can be found in Section 2. In Section \ref{sec:FOL}, we introduce first-order formulas for GP\,2 programs. In Section \ref{sec:SLP}, we outline the construction of a strongest liberal postcondition for a given rule schema and first-order formula. Section \ref{sec:proofrules} presents the proof rules of a semantic and a syntactic verification calculus, and identifies the class of programs that can be verified with the syntactic calculus.
In Section \ref{sec:ex}, we demonstrate how to verify a graph program for computing a 2-colouring of an input graph. In Section \ref{sec:completeness}, we discuss the soundness and completeness of our proof calculi. Then, in Section \ref{sec:related_work}, we compare our approach with other approaches in the literature. Finally, we conclude and give some topics for future work in Section \ref{sec:conclusion}.

\section{The Graph Programming Language GP\,2}
\label{sec:GP2}
In this section, we briefly review the graph programming language GP\,2 which was introduced in \cite{Plump12a}.

\subsection{GP\,2 Graphs} 
\label{sec:graphs}

A label in a GP\,2 graph consists of a list expression and an optional mark. The set $\mathbb{E}$ of expressions is defined by the grammar of \figurename~\ref{fig:Exp}. The set $\mathbb{L}$ of host graph lists is a subset of $\mathbb{E}$ and is defined by the grammar of \figurename~\ref{fig:List}.

\begin{figure}[ht]
\begin{subfigure}[b]{.5\textwidth}
\centering
\begin{scriptsize}
\begin{tabular}{lcl}%
      $\mathbb{E}$ & ::= &  List \\
      List & ::= & $\mtt{empty}$ $\mid$ Atom $\mid$ List \lq:' List $\mid$ ListVar \\
      Atom & ::= & Integer $\mid$ String $\mid$ AtomVar \\
      Integer & ::= & [\lq-']~Digit~\{Digit\}~$\mid$~\lq('Integer\lq)'  $\mid$ IntVar \\
       & & $\mid$ Integer (\lq+' $\mid$ \lq-' $\mid$ \lq*' $\mid$ \lq/') Integer \\
       & & $\mid$ ($\mtt{indeg}$ $\mid$ $\mtt{outdeg}$) \lq('NodeId\lq)' \\
       & & $\mid$ $\mtt{length}$ \lq('AtomVar $\mid$ StringVar $\mid$ ListVar\lq)' \\
      String & ::= & Char $\mid$ String `.' String $\mid$ StringVar\\
      Char & ::= & ` `` '\{Character\}` " ' $\mid$ CharVar\\
    \end{tabular}\end{scriptsize}
    \caption{Expressions (rule graph lists)}
    \label{fig:Exp}
\end{subfigure}
\begin{subfigure}[b]{.5\textwidth}
  \centering
  \begin{scriptsize}
\begin{tabular}{lcl}
      $\mathbb{L}$ & ::= & $\mtt{empty}$ $\mid$ GraphExp $\mid$ $\mathbb{L}$ `:' $\mathbb{L}$ \\
      GraphExp & ::= & [`-'] Digit \{Digit\} $\mid$ GraphStr\\
      GraphStr & ::= & `~``~' \{Character\} ' " ' $\mid$ GraphStr `.' GraphStr
    \end{tabular}\end{scriptsize}
 \caption{Host graph lists}
   \label{fig:List}
\end{subfigure}
\caption{Abstract syntax of GP\,2 lists}
\label{fig:GP2labels}
\end{figure}

Here Digit is the set $\{0,\ldots,9\}$ and Character is the set of all printable characters except `"' (i.e. the ASCII characters 32, 33, and 35-126). The variable sets ListVar, AtomVar, IntVar, StringVar, and CharVar contain variables of type $\mtt{list, atom,}$ $\mtt{int, string,}$ and $\mtt{char}$, respectively. The domains of \ttt{int} and \ttt{string} are the integers $\Z$ and the set $\mathrm{Character}^*$, respectively, while \ttt{atom} represents the union $\Z \cup \mathrm{Character}^*$.
The domain of \ttt{list} is $(\Z \cup \mathrm{Character}^*)^*$, the set of heterogeneous lists of integers and character strings. We identify lists and strings of length one with their contents and hence have the following subtype relationships: $\mtt{list} \supset \mtt{atom} \supset \mtt{string} \supset  \mtt{char}$ and $\mtt{atom} \supset \mtt{int}$.

The colon operator `:' is used to concatenate lists while the dot operator `.' is used to concatenate strings. The keyword $\mtt{empty}$ represents the empty list. The functions $\mtt{indeg}$ and $\mtt{outdeg}$ take a node as argument and return the indegree resp.\ outdegree of the node. The function $\mtt{length}$ takes a list or string variable as argument and returns the length of the list resp.\ string represented by the variable.

\begin{definition}[Rule graph]
\label{def:rulegraph}
\normalfont
Let $\mathbb{M}_V = \{\none,\red, \green, \blue, \grey\}$ be the set of \emph{node marks}\/ and $\mathbb{M}_E = \{\none,\red, \green, \blue, \dashed\}$ be the set of \emph{edge marks}. 

A \emph{rule graph} is a system $G = \langle V_G, E_G, s_G, t_G$, $l_G, m_G, p_G\rangle$ comprising a finite set $V_G$ of nodes, a finite set $E_G$ of edges, source and target functions $s_G, t_G\colon E_G \to V_G$, partial node labelling functions $\lst^V_G\colon V_G\to \mathbb{E}$ and $\mrk^V_G\colon V_G\to\mathbb{M}_V\cup\{\any\}$, edge labelling functions $\lst^E_G\colon E_G\to \mathbb{E}$ and $\mrk^E_G\colon E_G\to\mathbb{M}_E\cup\{\any\}$, and a partial root function $p_G\colon V_G \to \{0,1\}$. A rule graph is \emph{total}\/ if all of its functions are total functions. \hfill$\Box$
\end{definition}
\vspace{1ex}

The marks \ttt{red}, \ttt{green}, \ttt{blue} and \ttt{grey} are graphically represented by the obvious colours while \ttt{dashed} is represented by a dashed line. The wildcard mark $\mtt{any}$ is represented by the colour magenta. 

Node labels are undefined only in the interface graphs of rule schemata (see below). This allows rules to relabel nodes. Similarly, the root function is undefined only for the nodes of interface graphs. The purpose of root nodes is to speed up the matching of rule schemata \cite{Bak15a,Bak-Plump12a}.

Given a node $v$ in a graph $G$, we require that $\lst_G^V(v)$ is defined if and only if $\mrk_G^V(v)$ is defined. 

\begin{definition}[Host graph]
\label{def:hostgraph}
\normalfont
A \emph{host graph}\/ is a total rule graph $G$ satisfying $\lst_G^V(V_G) \subseteq \mathbb{L}$, $\lst_G^E(E_G) \subseteq \mathbb{L}$, $\mrk^V_G(V_G) \subseteq \mathbb{M}_V$ and $\mrk^E_G(E_G) \subseteq \mathbb{M}_E$.
\hfill$\Box$
\end{definition}
\vspace{1ex}

A \emph{graph morphism} $g:G\to H$ maps nodes to nodes and edges to edges such that sources, targets and labels are preserved. We also require that both roots and non-roots are preserved (see \cite{Campbell-Romo-Plump20a} for the root-reflecting mode of the GP\,2 compiler). A \emph{premorphism} is defined like a graph morphism except that labels need not be preserved.



\subsection{Conditional Rule Schemata} 
\label{sec:ruleschema}

The basic computational unit in GP\,2 are graph transformation rules labelled with expressions from $\mathbb{E}$, so-called rule schemata. They allow to modify the structure of host graphs and to perform computations on labels, such as arithmetic or list manipulations. Rule schemata can be equipped with application conditions to increase their expressiveness.

\Def{Conditional rule schema}{def:RS}{
A \emph{rule schema} $r=\langle L\leftarrow K\rightarrow R\rangle$ consists of two total rule graphs $L$ and $R$, and inclusion morphisms $K\to L$ and $K\to R$. Graph $K$ is the \emph{interface} of $r$ and consists of nodes only, with labels and roots  undefined. All expressions in $L$ must be \emph{simple}, that is, they do not contain arithmetic operators, contain at most one occurrence of a list variable, and contain at most one occurrence of a string variable in each occurrence of a string subexpression. Moreover, all variables in $R$ must also occur in $L$. A \emph{conditional rule schema} $\tuple{r,\, \Gamma}$ consists of a rule schema $r$ and an application condition $\Gamma$ according to the grammar of \figurename~\ref{fig:rule_schema_condition}, where all variables occurring in $\Gamma$ also occur in the left-hand graph of $r$.}

\begin{figure}[hbt]
\small
\begin{center}
\begin{tabular}{lcl}
        Condition & ::= & ($\mtt{int \mid char \mid string \mid atom}$) `('Var`)' \\
        && $\mid$ List (`\ttt{=}' $\mid$ `\ttt{!=}') List $\mid$ Integer (`\ttt{>}' $\mid$ `\ttt{>=}' $\mid$ `\ttt{<}' $\mid$ `\ttt{<=}') Integer\\
        && $\mid$ $\mtt{edge}$ `(' NodeId `,' NodeId [`,' List [EdgeMark]] `)'\\
        && $\mid$ $\mtt{not}$ Condition $\mid$ Condition ($\mtt{and}$ $\mid$ $\mtt{or}$) Condition $\mid$ `(' Condition `)'\\
        Var & ::= & ListVar $\mid$ AtomVar $\mid$ IntVar $\mid$ StringVar $\mid$ CharVar\\
        EdgeMark & ::= & $\mtt{red \mid green \mid blue \mid dashed \mid any}$
\end{tabular}
\end{center}
    \caption{Application conditions for rule schemata}
    \label{fig:rule_schema_condition}
\end{figure}

A conditional rule schema $\tuple{L\gets K\to R,\, \Gamma}$ is applied to a host graph $G$ in stages: (1) evaluate the expressions in $L$ and $R$ with respect to a premorphism $g\colon L \to G$ and a label assignment $\alpha$, obtaining an instantiated rule $\tuple{L^{g,\alpha}\gets K\to R^{g,\alpha}}$; (2) check that $g\colon L^{g,\alpha} \to G$ is label preserving and that the evaluation of $\Gamma$ with respect to $g$ and $\alpha$ returns true; (3) construct two natural pushouts based on the instantiated rule and $g$.

\Def{Label assignment}{def:assign}{
Consider a rule graph $L$ and the set $X$ of all variables occurring in $L$. For each $x\in X$, let dom$(x)$ denote the domain of $x$ associated with the type of $x$. A \emph{label assignment} for $L$ is a triple $\alpha=\tuple{\alpha_\mathbb{L},\, \mu_V,\, \mu_E}$ where $\alpha_\mathbb{L}\colon X\rightarrow\mathbb{L}$ is a function such that for each $x\in X$, $\alpha_\mathbb{L}(x)\in$\,dom$(x)$, and $\mu_V\colon V_L\to \mathbb{M}_V\backslash\{\mtt{none}\}$ and $\mu_E\colon E_L\to \mathbb{M}_E\backslash\{\mtt{none}\}$ are partial functions assigning a mark to each node and edge marked with \ttt{any}.
}

Given a rule graph $M$, a host graph $G$, an injective premorphism $g\colon M \to G$, and a label assignment $\alpha=\tuple{\alpha_\mathbb{L},\, \mu_V,\, \mu_E}$ for $M$, the \emph{instance} $M^{g,\alpha}$ is obtained as follows: (1) replace each variable $x$ in a list expression with $\alpha_\mathbb{L}(x)$; (2) replace each \ttt{any} mark of a node $v$ or edge $e$ with $\mu_V(v)$ resp.\ $\mu_E(e)$; (3) replace each node identifier $n$ in a list expression with $g(n)$; (4) evaluate all resulting list expressions according to the meaning of the operators in Figure \ref{fig:Exp} (see \cite{Bak15a} for details). Note that $M^{g,\alpha}$ is a host graph.

The instance $\Gamma^{g,\alpha}$ of an application condition $\Gamma$ is obtained by applying steps (1) and (3), and evaluating the resulting condition according to the meaning of the operators in Figure \ref{fig:rule_schema_condition} (see \cite{Bak15a} for details). Note that $\Gamma^{g,\alpha}$ is either ``true" or ``false".

\begin{definition}[Conditional rule schema application]
\label{def:ruleapp}
\normalfont
Consider a conditional rule schema $r=\lkr$, host graphs $G$ and $H$, and an injective premorphism $g\colon L \to G$. Then $G$ \emph{directly derives} $H$ by $r$ and $g$, denoted by $G \dder_{r,g} H$, if there exists a label assignment $\alpha$ for $L$ such that
\vspace{-\topsep}\begin{enumerate}\setlength{\parskip}{0pt} \setlength{\itemsep}{0pt plus 1pt}
    \item[(i)] $g\colon L^{g,\alpha} \to G$ is a label preserving graph morphism, 
    \item[(ii)] $\Gamma^{g,\alpha}$ is true,
    \item[(iii)] $G \dder_{r^{g,\alpha},g} H$.
\end{enumerate}\vspace{-\topsep}
Here $G \Rightarrow_{r^{g,\alpha},g} H$ denotes the existence of the following natural double-pushout:\footnote{A pushout is \emph{natural} if it is also a pullback.}

$~~~~~~~~~~~~~~~~~~~~~~~~~~~~~~~~~~~~~~~~~~~~~~~~~~~~~~~~~$
\begin{tikzpicture}[scale=0.8, transform shape]
	\node (2) at (0.75, 0) {};
	\node (a) at (1.5, 0.75) {$K$};
	\node (b) at (0, 0.75) {$L^\alpha$};
	\node (c) at (3, 0.75) {$R^\alpha$};
	\node (d) at (1.5, -0.75) {$D$};
	\node (1) at (2.25, 0) {};
	\node (e) at (3, -0.75) {$H$};
	\node (f) at (0, -0.75) {$G$};
	\draw[->] (a) to node {} (b);
	\draw[->] (a) to node {} (c);
	\draw[->] (a) to node {} (d);
	\draw[->] (d) to node {} (e);
	\draw[->] (d) to node {} (f);
	\draw[->] (c) to node [right]{$g^*$} (e);
	\draw[->] (b) to node [left]{$g$} (f);
    \end{tikzpicture} \hfill$\square$ 
\end{definition}

Given $r$ and $g$ such that (i) and (ii) are satisfied, there exists a natural double-pushout as above if and only if $g$ satisfies the \emph{dangling condition}: no node in $g(L-K)$ must be incident to an edge in $G - g(L)$. 

In graph transformations, usually a derivation do not require the double-pushouts to be natural \cite{Ehrig06}. Here, we require them to be natural due to relabelling (see \cite{Bak15a,Campbell-Romo-Plump20a} for the motivation of using natural double-pushouts and for their construction).

A rule schema $r$ without application condition can be considered as the conditional rule schema $\tuple{r,\,\Delta}$ where $\Delta$ is a condition that is always true (such as \ttt{0{=}0}). In this case, point (ii) in the above definition is trivially satisfied.

\subsection{Syntax and Semantics of Programs} 
\label{sec:commands}

 A graph program consists of declarations of conditional rule schemata and procedures, and exactly one declaration of a main command sequence, which is a distinct procedure named \texttt{Main}. Procedures must be non-recursive, they can be seen as macros. The syntax of GP\,2 programs is defined by the grammar in Figure \ref{fig:GP2syntax} (where we omit the syntax of rule schema declarations). In the following we describe the main control constructs.
    
\begin{figure}[htb]
\footnotesize 
\begin{center}
\begin{tabular}{lll}
Prog & ::= & Decl \{Decl\}\\
Decl & ::= & MainDecl $\mid$ ProcDecl $\mid$ RuleDecl\\
MainDecl & ::= & $\mtt{Main}$ `=' ComSeq\\
ProcDecl & ::= & ProcId `=' Comseq\\
ComSeq & ::= & Com \{`;' Com\}\\
Com & ::= & RuleSet $\mid$ Proc\\
&& $\mid~\mtt{if}$ ComSeq $\mtt{ then }$ ComSeq [$\mtt{else }$ ComSeq]\\
&& $\mid~\mtt{try }$ ComSeq [$\mtt{ then }$ ComSeq] [$\mtt{else }$ ComSeq]\\
&& $\mid~$ComSeq `!' $\mid~$ComSeq $\mtt{ or }$ ComSeq $\mid~$`(' ComSeq `)'\\
&& $\mid~\mtt{break}~\mid~\mtt{skip}~\mid~\mtt{fail}$\\
RuleSet & ::= & RuleId $\mid$ `\{' [RuleId \{ `,' RuleId\}] `\}'\\
Proc & ::= & ProcId
\end{tabular}
\end{center}
\caption{Abstract syntax of GP\,2 programs}
\label{fig:GP2syntax}
\end{figure}

The call of a rule set $\{r_1,\dots,r_n\}$ non-deterministically applies one of the rules whose left-hand graph matches a subgraph of the host graph such that the dangling condition and the rule's application condition are satisfied. The call \emph{fails}\/ if none of the rules is applicable to the host graph. 
    
The command \ttt{if} $C$ \ttt{then} $P$ \ttt{else} $Q$ is executed on a host graph $G$ by first executing $C$ on a copy of $G$. If this results in a graph, $P$\/ is executed on the original graph $G$; otherwise, if $C$ fails, $Q$ is executed on $G$. The \ttt{try} command has a similar effect, except that $P$\/ is executed on the result of $C$'s execution. 
    
The loop command $P!$ executes the body $P$\/ repeatedly until it fails. When this is the case, $P!$ terminates with the graph on which the body was entered for the last time. The \ttt{break} command inside a loop terminates that loop and transfers control to the command following the loop.

\begin{figure}[h!]
\centering
\scalebox{0.9}{
\begin{small}
\begin{tabular}{lll}
~[Call$_1$]$\displaystyle\frac{G\dder_{\R} H}{\langle \R,\, G\rangle \to H}$&$~~~~~~~$&[Call$_2$]$\displaystyle\frac{G\nRightarrow_{\R}}{\langle \R,\, G \rangle \to \mrm{fail}}$\\~\\
~[Seq$_1$]$\displaystyle\frac{\langle P,\, G\rangle\rightarrow\langle P',\, H\rangle}{\langle P;Q,\, G\rangle\rightarrow\langle P';Q,\, H\rangle}$&&[Seq$_2$]$\displaystyle\frac{\langle P,\, G\rangle\rightarrow H}{\langle P;Q,\, G\rangle\rightarrow\langle Q,\, H\rangle}$\\~\\
~[Seq$_3$]$\displaystyle\frac{\langle P,\,G\rangle\rightarrow\mrm{fail}}{\langle P;Q,\, G\rangle\rightarrow\mrm{fail}}$&&[Break]$\displaystyle\frac{}{\langle \texttt{break};P,\, G\rangle\rightarrow\langle \texttt{break},\, G\rangle}$\\~\\
~[If$_1$]$\displaystyle\frac{\langle C,\, G\rangle\rightarrow ^+ H}{\langle \texttt{if }C\texttt{ then }P\texttt{ else }Q,\, G\rangle\rightarrow\langle P,\, G\rangle}$&&[If$_2$]$\displaystyle\frac{\langle C,\, G\rangle\rightarrow ^+ \mrm{fail}}{\langle \texttt{if }C\texttt{ then }P\texttt{ else }Q,\, G\rangle\rightarrow\langle Q,\, G\rangle}$\\~\\
~[Try$_1$]$\displaystyle\frac{\langle C,\, G\rangle\rightarrow ^+ H}{\langle \texttt{try }C\texttt{ then }P\texttt{ else }Q,\, G\rangle\rightarrow\langle P,\, H\rangle}$&&[Try$_2$]$\displaystyle\frac{\langle C,G\rangle\rightarrow ^+ \mrm{fail}}{\langle \texttt{try }C\texttt{ then }P\texttt{ else }Q,\, G\rangle\rightarrow\langle Q,\, G\rangle}$\\~\\
~[Loop$_1$]$\displaystyle\frac{\langle P,\, G\rangle\rightarrow ^+H}{\langle P!,\, G\rangle\rightarrow\langle P!,\, H\rangle}$&&[Loop$_2$]$\displaystyle\frac{\langle P,\, G\rangle\rightarrow ^+\mrm{fail}}{\langle P!,\, G\rangle\rightarrow G}$\\~\\
~[Loop$_3$]$\displaystyle\frac{\langle P,\, G\rangle\rightarrow ^*\langle\texttt{break}, \, H\rangle}{\langle P!,\, G\rangle\rightarrow H}$&&
\end{tabular}\end{small}}
\caption{Semantic inference rules for GP\,2 core commands}
\label{fig:infRule-core}
\end{figure}

In general, the execution of a program on a host graph may result in different graphs, fail, or diverge. The operational semantics of GP\,2 is defined by the inference rules of Figure \ref{fig:infRule-core}, where $\R$ stands for a rule set call; $C,P,P'$, and $Q$ stand for command sequences; and $G$ and $H$ stand for host graphs. Given a program $P$, the rules induce a semantic function which maps each host graph $G$ to the set $\llbracket P\rrbracket G$ of all possible outcomes of executing $P$ on $G$. The result set may contain proper results in the form of graphs and the special values ``fail" and $\perp$. The value ``fail" indicates a failed program run while $\perp$ indicates a run that diverges. Hence the set of all configurations is  $(\mrm{ComSeq} \times \G(\mathbb{L})) \cup \G(\mathbb{L}) \cup \{\mrm{fail}\}$, where ComSeq is the set of command sequences as defined in Figure \ref{fig:GP2syntax} and $\mathcal{G}(\mathbb{L})$ is the set of all host graphs.



\section{First-Order Formulas for Graph Programs}
\label{sec:FOL}
In this section, we define first-order formulas which specify classes of GP\,2 graphs. We also show how to represent concrete GP\,2 graphs in rule schema applications. 

\subsection{Syntax of First-Order Formulas}
To be able to express GP\,2 graphs, we need to be able to express properties of a graph and GP\,2 rule schema conditions. Here, we only consider totally labelled graphs. Lists in GP\,2 graphs can be expressed by variables. In our first-order formulas, variables may express nodes or edges as well (see Table \ref{tab:domkind}).

\begin{table}[h!]
    \centering
    \caption{Kind of a variable and its domain in a graph $G$}\label{tab:domkind}
     \begin{footnotesize}\begin{tabular}{|l|c|c|c|c|c|c|c|} 
    \hline
        \textbf{kind of variables} & Node & Edge & List & Atom & Int & String & Character\\\hline
        \textbf{domain} 
         & $V_G$ & $E_G$ &
         $(\mathbb{Z}\cup(\text{Char})^*)^*$ &
         $\mathbb{Z}\cup\text{Char}^*$ &
         $\mathbb{Z}$ &
        $\text{Char}^*$ &
         Char \\\hline

    \end{tabular}
    \end{footnotesize}
\end{table}

The syntax of first-order (FO) formulas is given by the grammar of Figure \ref{fig:mso}. In the syntax, NodeVar and EdgeVar represent disjoint sets of first-order node and edge variables, respectively. We use ListVar, AtomVar, IntVar, StringVar, and CharVar for sets of first-order label variables of type $\mrm{list, atom, int, string}$, and $\mrm{char}$ respectively. The nonterminals Character and Digit in the syntax represent the fixed character set of GP\,2 characters, and the digit set $\{0,\ldots,9\}$ respectively, as what we have in the syntax of \figurename~\ref{fig:GP2labels}.

\begin{figure}
    \centering
    \begin{small}
    \begin{tabular}{lcl}
      Formula & ::= & $\mrm{true}~\mid~\mrm{false}~\mid$ Cond $\mid$ Equal\\
        && $\mid$ Formula~(`$\mrm{\wedge}$' $\mid$ `$\mrm{\vee}$')~Formula $\mid$ `$\neg$'Formula $\mid$ `('Formula`)'\\
        && $\mid$ `$\exists_\mathtt{V}$' (NodeVar)~`('Formula`)' \\
        && $\mid `\exists_\mathtt{E}$'~(EdgeVar)~`('Formula`)'       \\
        && $\mid$ `$\exists_\mathtt{L}$' (ListVar)~`('Formula`)' \\
      Number & ::= & Digit~\{Digit\}\\
      Cond & ::= & ($\mrm{int \mid char \mid string \mid atom}$) `('Var`)' \\
        && $\mid$ Lst (`$\mrm{=}$' $\mid$ `$\mrm{\neq}$') Lst $\mid$ Int (`$\mrm{>}$' $\mid$ `$\mrm{>=}$' $\mid$ `$\mrm{<}$' $\mid$ `$\mrm{<=}$') Int\\
        && $\mid$ $\mrm{edge}$ `(' Node `,' Node [`,' Lst] [`,' EMark] `)' $\mid$ $\mrm{root}$ `(' Node `)'\\
      Var & ::= & ListVar $\mid$ AtomVar $\mid$ IntVar $\mid$ StringVar $\mid$ CharVar\\
      Lst & ::= & $\mrm{empty}$ $\mid$ Atm $\mid$ Lst \lq:' Lst $\mid$ ListVar $\mid$ $\mrm{l_V}$ `('Node`)' $\mid$ $\mrm{l_E}$ `('EdgeVar`)'  \\
      Atm & ::= & Int $\mid$ String $\mid$ AtomVar \\
      Int & ::= & [\lq-']~Number~$\mid$~\lq('Int\lq)'  $\mid$ IntVar $\mid$ Int (\lq+' $\mid$ \lq-' $\mid$ \lq*' $\mid$ \lq/') Int \\
        && $\mid$ ($\mrm{indeg}$ $\mid$ $\mrm{outdeg}$) \lq('Node\lq)' $\mid$ $\mrm{length}$ \lq('AtomVar $\mid$ StringVar $\mid$ ListVar\lq)' \\
      String & ::= & ` `` ' {Character} ` " ' $\mid$ CharVar $\mid$ StringVar $\mid$ String `.' String \\
      Node & ::= & NodeVar $\mid$ ($\mrm{s}~\mid \mrm{t}$) `(' EdgeVar`)'\\
      EMark & ::= & $\mrm{none~\mid~red~\mid~green~\mid~blue~\mid~dashed~\mid~any}$\\
      VMark & ::= & $\mrm{none~\mid~red~\mid~blue~\mid~green~\mid~grey~\mid~any}$  \\
      Equal & ::= & Node ('$\mrm{=}$' $\mid$ `$\mrm{\neq}$') Node $\mid$ EdgeVar ('$\mrm{=}$' $\mid$ `$\mrm{\neq}$') EdgeVar\\
        && $\mid$ Lst ('$\mrm{=}$' $\mid$ `$\mrm{\neq}$') Lst $\mid$ $\mrm{m_V}$`('Node`)' ('$\mrm{=}$' $\mid$ `$\mrm{\neq}$') VMark \\
        && $\mid$ $\mrm{m_E}$`('EdgeVar`)' ('$\mrm{=}$' $\mid$ `$\mrm{\neq}$') EMark
    \end{tabular}\end{small}
    \caption{Syntax of first-order formulas}
    \label{fig:mso}
\end{figure}

The quantifiers $\E{V}, \E{E},$ and $\E{L}$ in the grammar are reserved for variables of nodes, edges, and labels respectively. The function symbols $\mrm{indeg, outdeg}$ and $\mrm{length}$ return indegree, outdegree, and length of the given argument. Also, we have unary functions $\mrm{s, t, l_V, l_E, m_V,}$ and $\mrm{m_{E}}$, which takes the argument and respectively return the value of its source, target, node label, edge label, node mark, and edge mark. The predicate $\mrm{edge}$ expresses the existence of an edge between two nodes. The predicates $\mrm{int, char, string, atom}$ are typing predicates to specify the type of the variable in their argument. When a variable is not an argument of any typing predicate, then the variable is a list variable. We have the predicate $\mrm{root}$ to express rootedness of a node. For brevity, we sometimes write $\mrm{\forall_Vx}(c)$ for $\neg\E{V}x(\neg c)$ and $\mrm{\E{V}x_1,\ldots,x_n}(c)$ for $\mrm{\E{V}x_1(\E{V}x_2(...\E{V}x_n}(c)\ldots))$ (also for edge and label quantifiers). Also, we define 'terms' as the set of variables, constants, and functions in first-order formulas.

The satisfaction of a FO formula $c$ in a host graph $G$ relies on \textit{assignments}. An assignment $\alpha$ of a formula $c$ on $G$ is a pair $\tuple{\alpha_G,\alpha_\mathbb{L}}$ where $\alpha_G$ is function that maps every free node (or edge) variable to a node (or edge) in $G$, and $\alpha_\mathbb{L}$ is a function that maps every free char, string, integer, atom, and list variable in $c$ to a member of its domain based on Table \ref{tab:domkind}. From an assignment $\alpha$, we can obtain $c^\alpha$ by replacing every free variable $x$ with $\alpha(x)$, and evaluate the functions based on the semantics of their associated GP\,2 syntax. $G$ satisfies $c$ by assignment $\alpha$, denotes by $G\Sata c$ if and only if $c^\alpha$ is true in $G$. 

The truth value of $c^\alpha$ is evaluated just like in standard logic, with respect to the semantics of the predicates as described above, where $\mrm{(root(x))^\alpha}$ is true in $G$ if $x^\alpha$ is rooted, or false otherwise. We then write $G\Sat c$ if there exists an assignment $\alpha$ such that $G\Sata c$.

\subsection{Conditions for Rule Schema Applications}
First-order formulas as defined above do not contain node or edge constants because we want to be able to check the satisfaction of formulas on arbitrary host graphs. However, for rule schema applications we will need to express properties of specific nodes and edges of the graphs in the rule schema. For this, we define a \textit{condition over a graph} that can be obtained from a first-order formula and an assignment. 

\Def{Conditions}{def:condition}{
A \textit{condition} is a first-order formula without free node and edge variables. A \textit{condition over a graph} $G$ is a first-order formula where every free node and edge variable is replaced with node and edge identifiers in $G$. That is, if $c$ is a FO formula and $\alpha_G$ is an assignment of free node and edge variables of $c$ on $G$, then $c^{\alpha_G}$ is a condition over $G$.}

Checking if a graph satisfies a condition $c$ over a graph is essentially similar to checking satisfaction of a FO formula in a graph. However, the satisfaction of $c$ in a graph $G$ can be defined only if $c$ is a condition over $G$. 

Given a rule schema $\tuple{L\leftarrow K\rightarrow R}$ and an injective morphism $g:L\rightarrow G$ for some host graph $G$. The satisfaction of a condition $c$ over $L$ may not be defined in $G$. However, we can rename some nodes and edges in $G$ with respect to $g$ so that $c$ is a condition over the graph (with renamed nodes and edges).

\Def{Replacement graph}{def:rho}{
Given an injective morphism $g:L\rightarrow G$ for host graphs $L$ and $G$. Graph $\RG$ is a replacement graph of $G$ w.r.t. $g$ if $\RG$ is isomorphic to $G$ with $L$ as a subgraph.}

A conditional rule schema is not invertible because of the restrictions on the variables and the existence of the rule schema condition that is reserved only for the left-hand graph. However, an invertible rule is sometimes needed to be able to derive properties from output graph to the input graph. Hence, we define a generalisation of a rule schema. Here, we define an \emph{unrestricted rule schema} as a rule schema without any restriction on the occurring labels. 

\Def{Generalised rule}{def:genrule}{
Given an unrestricted rule schema $r=\langle L\leftarrow K\rightarrow R\rangle$.
\emph{A generalised rule} is a tuple $w=\langle r,ac_L,ac_R\rangle$ where $ac_L$ is a condition over $L$ and $ac_R$ is a condition over $R$. We call $ac_L$ the left application condition and $ac_R$ the right application condition. The inverse of $w$, written $w^{-1}$, is then defined as the tuple $\langle r^{-1},ac_R,ac_L\rangle$ where $r^{-1}=\langle R\leftarrow K\rightarrow L\rangle$.}

The application of a generalised rule is essentially similar to the application of a rule schema. However in a generalised version, we need to consider the satisfaction of both left and right-application condition in the replacement graph of input and output graphs. For a conditional rule schema $r=\tuple{\tuple{L\leftarrow K\rightarrow R},\Gamma}$, we denote by $r^\vee$ the general version of $r$, that is the generalised rule $r^\vee=\tuple{\tuple{L\leftarrow K\rightarrow R},\Gamma^\vee,\mrm{true}}$ where $\Gamma^\vee$ is obtained from $\Gamma$ by replacing the notations !=, $\mtt{not},\mtt{and, or}, \#$ with $\neq, \neg, \wedge, \vee, `,$\rq (comma symbol) respectively.

\section{Constructing a Strongest Liberal Postcondition}
\label{sec:SLP}
In this section, we show how to construct a strongest liberal postcondition from a given conditional rule schema and a precondition. The condition expresses properties that must be satisfied by every graph resulting from the application of the rule schema to a graph satisfying the given precondition \cite{DijkstraS90}. Here, a precondition is limited to a closed FO formula.

\Def{Strongest Liberal Postcondition}{def:slpP}{
An assertion $d$ is a \emph{liberal postcondition} with respect to a precondition $c$ and a graph program $P$, if for all host graphs $G$ and $H$,\\
$~~~~~~~~~~~~~~~~~~~~~~~~~~~~~~~~~~~~~~~~~~~~~~(G\vDash c \text{ and } H\in\Sem{P}G) \text{ implies }H\vDash d.$\\
A \emph{strongest liberal postcondition} w.r.t.\ $c$ and $P$, denoted by SLP$(c,P)$, is a liberal postcondition w.r.t.\ $c$ and $P$ that implies every liberal postcondition w.r.t.\ $c$ and $P$.}

To construct $\text{SLP}(c,r)$, we use the generalised version of $r$ to open a possibility of constructing a strongest liberal postcondition over the inverse of a rule schema. $\text{SLP}(c,r)$ is obtained by defining transformations Lift$(c,r^\vee)$, Shift$(c,r^\vee)$, and Post$(c,r^\vee)$. The transformation Lift transforms the given condition $c$ into a left-application condition w.r.t. $r^\vee$, which is then transformed into a right-application condition by Shift. Finally, the transformation Post transforms the right-application condition to $\text{SLP}(c,r)$. Similar approach has been used in \cite{Poskitt13,HP09,Pennemann09} for constructing a weakest liberal precondition from a given postcondition. 

\begin{figure}[h]
\centering  
  \begin{tikzpicture}[remember picture,
  inner/.style={circle,draw,minimum size=18pt},
  outer/.style={inner sep=2pt}, scale=0.6
  ]
  \node[outer] (AA) at (0,1) {\footnotesize{$\mtt{del(a,b,c:list;~ d,e:int)}$}};
  \node[outer] (A) at (0,0) {
  \begin{tikzpicture}[scale=0.6, transform shape]
		\node[inner, label=below:\tiny 1] (Aa) at (0,0) {$\mtt{a}$};	
		\node[inner, label=below:\tiny 2] (Ab) at (1.5,0) {$\mtt{b}$};	
        \node[inner, label=below:\tiny 3] (Ac) at (-1.5,0) {$\mtt{c}$};	
		\draw[-latex] (Aa) to node[above] {$\mtt{d}$} (Ab);
		\draw[-latex] (Aa) to node[above] {$\mtt{e}$} (Ac);
		\end{tikzpicture}};
  \node[outer] (B) at (0,-1) {\footnotesize{$\mtt{where~d\geq e}$}};
  \node[outer] (B) at (2.25,0) {$\Rightarrow$};
  \node[outer] (C) at (3.75,0) {
  \begin{tikzpicture}[scale=0.6, transform shape]
		\node[inner, label=below:\tiny 1, fill=red!50] (Aa) at (0,0) {$\mtt{a}$};	
		\node[inner, label=below:\tiny 2] (Ab) at (1.75,0) {$\mtt{b}$};	
        \draw[-latex] (Aa) to node[above] {$\mtt{d+e}$} (Ab);
		\end{tikzpicture}};
	\end{tikzpicture}
\caption{GP\,2 conditional rule schema $\mtt{del}$}
\label{fig:ruledel}
\end{figure}
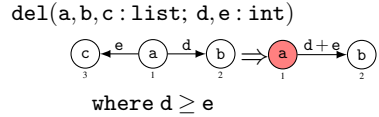

To give a better idea of the transformations we define in this section, we show a running example for the construction. We use the conditional rule schema $\mtt{del}$ of \figurename~\ref{fig:ruledel} and the preconditions $q=\mrm{\neg\E{E}x(m_V(s(x))\neq none)}$ for the running example. We denote by $\Gamma_1$ the GP\,2 rule schema condition $\mrm{d\geq e}$. In addition, a simple example of the construction can be seen in Section \ref{sec:ex}. 

\subsection{From Precondition to Left-Application Condition}

Now, we start with transforming a precondition $c$ to a left-application condition with respect to a generalised rule $w=\tuple{r,ac_L,ac_R}$. Intuitively, the transformation is done by:
\vspace{-\topsep}\begin{enumerate}
    \item Find all possibilities of variables in $c$ representing nodes/edges in an input and form a disjunction from all possibilities, denoted by Split$(c,r)$;
    \item Express the dangling condition as a condition over $L$, denoted by Dang$(r)$;
    \item Evaluate terms and Boolean expression in Split$(c,r)$, Dang$(r)$, and $\Gamma^\vee$, then form a conjunction from the result of evaluation, and simplify the conjunction.
\end{enumerate}

A possibility of variables in $c$ representing nodes/edges in an input graph as mentioned above refers to a way variables in $c$ can represent node or edge constants in the replacement of the input graph. A simple example would be for a precondition $c=\E{V}x(c_1)$ for some FO formula $c_1$ with a free variable $x$, $c$ holds on a host graph $G$ if there exists a node $v$ in $G$ such that $c_1^{\alpha}$ where $\alpha(x)=v$ is true in $G$. In the replacement graph of $G$, $v$ can be any node in the left-hand graph of the rule schema, or any node outside it. Split$(c,r)$ is obtained from the disjunction of all these possibilities.


\Def{Transformation Split}{def:split}{
Given an unrestricted rule schema $r=\tuple{L\leftarrow K\rightarrow R}$. where $V_L=\{v_1,\ldots,v_n\}$ and $E_L=\{e_1,\ldots,e_m\}$. Let $c$ be a condition over $L$ sharing no variables with $r$ (note that it is always possible to replace the label variables in $c$ with new variables that are distinct from variables in $r$). We define the condition $\text{Split}(c,r)$ over $L$ inductively as follows:\\
\begin{small}
\begin{tabular}{ll}
    \multicolumn{2}{l}{- Base case.}  \\
     $~$ &  If $c$ is $\mrm{true}$, $\mrm{false}$, a predicate $\mrm{int(t), char(t), string(t), atom(t), root(t)}$ for \\&some term $\mrm{t}$, or in the form $\mrm{t_1\ominus t_2}$ for $\mrm{\ominus\in\{=.\neq.<,\leq,>,\geq\}}$ and some\\& terms $\mrm{t_1, t_2}$,\\
     & \multicolumn{1}{c}{\text{Split$(c,r) = c$}}\\
    \multicolumn{2}{l}{- Inductive case.}  \\
     & Let $c_1$ and $c_2$ be conditions over $L$.\\
     & 1) $\text{Split}(c_1\vee c_2, r) = \text{Split}(c_1, r)\vee\text{Split}(c_2, r)$,\\
    & 2) $\text{Split}(c_1 \wedge c_2, r) = \text{Split}(c_1, r)\wedge\text{Split}(c_2, r)$,\\
    & 3) $\text{Split}(\neg c_1, r) = \neg\text{Split}(c_1, r)$,\\
    & 4) $\text{Split}(\mrm{\E{V}x}(c_1), r)= (\mrm{\bigvee_{i=1}^n}\text{Split}(c_1^{[ x\mapsto v_i]}, r))\vee\mrm{\E{V} x(\bigwedge_{i=1}^n x{\neq}v_i\,\wedge\,} \text{Split}(c_1, r)$,\\
    & 5) $\text{Split}(\mrm{\E{E}x}(c_1), r)=\mrm{(\bigvee_{i=1}^m}\text{Split}(c_1^{[x\mapsto e_i]}, r))\vee\mrm{\E{E}x(\bigwedge_{i=1}^m x{\neq}e_i\,\wedge\,} \text{inc}(c_1, r,x))$,\\
   & ~~~~where\\
   & $~~~~\text{inc}(c_1, r,x)=\mrm{\bigvee_{i=1}^n (\bigvee_{j=1}^n s(x)=v_i\wedge t(x)=v_j\,\wedge\,}\text{Split}(c_1^{[\mrm{s(x)\mapsto v_i, t(x)\mapsto v_j}]}, r))$\\
   & $~~~~~~~~~~~~~~~~~~~~~~~~~~~\mrm{\vee\,
   (s(x)=v_i\,\wedge\,\bigwedge_{j=1}^n t(x)\neq v_j\,\wedge\,}\text{Split}(c_1^{[\mrm{s(x)\mapsto v_i}]}, r))$\\
   & $~~~~~~~~~~~~~~~~~~~~~~~~~~~\mrm{\vee\,
   (\bigwedge_{j=1}^n s(x)\neq v_j\,\wedge\,t(x)= v_i\,\wedge\,}\text{Split}(c_1^{[\mrm{t(x)\mapsto v_i}]}, r))$\\
   & $~~~~~~~~~~~~~~~~~~~~~\mrm{\vee\,
   (\bigwedge_{i=1}^n s(x)\neq v_i\,\wedge\,\bigwedge_{j=1}^n t(x)\neq v_j\,\wedge\,}\text{Split}(c_1
   , r))$\\
   & 6) $\text{Split}(\mrm{\E{L}x}(c_1), r)=\E{L}\mrm{x}(\text{Split}(c_1, r))$
\end{tabular}\\
\noindent where $c^{[a\mapsto b]}$ for a variable $a$ and constant $b$ represents the condition $c$ after the replacement of all occurrence of $a$ with $b$. Similarly, $c^{[d\mapsto b]}$ for $d\in\{\mrm{s(x), t(x)}\}$ is also a replacement $d$ with $b$.
\end{small}
}

In constructing Split$(c,r)$, the replacement for an edge quantifier is not as simple as the replacement for a node quantifier. For an edge variable $x$ in a precondition, $x$ can represent any edge in $G$. Moreover, if the condition contains the term $\mrm{s(x)}$ or $\mrm{t(x)}$, it may represent a node in the image of the match. Hence, we need to check these possibilities as well.

\Ex{Transformation Split}{ex:Split}{$~$\\
    Split$(q, \mtt{del}) = \mrm{\neg(m_V(s(e1))\neq none \vee m_V(s(e2))\neq none}$\\
     $\mrm{~~~~~~~~~~~~~~~~~~~~~~~~~~~~~\vee\,\E{E}x(x\neq e1\wedge x\neq e2\,\wedge\,((s(x)=1\wedge m_V(1)\neq none)\vee\,(s(x)=2\wedge m_V(2)\neq none)}$\\
      $~~~~~~~~~~~~~~~~~~~~~~~~~~~~~~~~~~~~~~~~~~~~~~~~~~~~~~~~~~~~~~~~~~~~~~~\mrm{\vee\,(s(x)=3\wedge m_V(3)\neq none)}$\\
      $~~~~~~~~~~~~~~~~~~~~~~~~~~~~~~~~~~~~~~~~~~~~~~~~~~~~~~~~~~~~~~~~~~~~~~~\mrm{\vee\,(s(x)\neq 1\wedge s(x)\neq 2\wedge s(x)\neq 3\wedge m_V(s(x))\neq none))))}$\\
}

Besides obtaining Split$(c,r)$, we also need to express the dangling condition as a condition over $L$. The dangling condition must be satisfied by an injective morphism $g$ if $G\Rightarrow_{r,g}H$ for some rule schema $r=\tuple{L\leftarrow K\rightarrow R}$ and host graphs $G,H$. Since we want to express properties of $\RG$ where such derivation exists, we need to express the dangling condition as a condition over the left-hand graph. For every node $v\in L-K$, the dangling condition is satisfied if and only if $v$ is not incident to any edge not in $L$. Therefore, the indegree and outdegree of $v$ in $\RG$ must be equal to the indegree and outdegree of $v$ in $L$. Hence, if we have $V_L-V_K=\{v_1,\ldots,v_n\}$, we can have:\\
(i) Dang($r)=\mrm{true}$ if $V_L-V_K=\emptyset$, and\\ (ii) Dang$(r)=\mrm{\bigwedge_{i=1}^n indeg(v_i)=}indeg_L(v_i)\,\wedge\,\mrm{outdeg(v_i)=}outdeg_L(v_i)$ otherwise.

\Ex{Dangling Condition}{ex:dang}{
Dang$(\mtt{del}) = \mrm{indeg(3)=1\wedge outdeg(3)=0}$
}

Since we have information about some properties of $L$ from the rule, we can put the information in the condition by evaluating the condition we obtained from Split and Dang with respect to $L$. For this, we construct of Val($d,r$) for a condition $d$ over $L$ where $L$ is the left-hand graph of $r$. Intuitively, Val($d,r$) is obtained from $d$ by replacing every term with its value in $L$ where possible. Possible here means if the argument of the term contains a constant. We then simplify the resulting condition so that there is no subformula in the form $\mrm{\neg\, true,} \neg(\neg\,a)$ ${\neg(a\vee b),}$ ${\neg(a\wedge b)}$ for some conditions $a, b$. We can simplify them to $\mrm{false}, a, \neg a\wedge\neg b, \neg a\vee\neg b$ respectively.

There is a special case when the term is in the form $\mrm{indeg(x)}$ or $\mrm{outdeg(x)}$ because unlike the other terms, their value in $L$ is different with their value in the replacement graph of the input graph. For more information about handling this case, we refer readers to \cite{WP20a}.

\begin{example}[Valuation of a Graph Condition]\label{ex:val}\normalfont $~$
\vspace{-\topsep}\begin{enumerate}
    \item \begin{tabular}[t]{lcl}
    \multicolumn{3}{l}{Val$(\text{Split}(q, \mtt{del}), \mtt{del})$}\\
    & = & $\mrm{\neg(none\neq none \vee none\neq none}$\\
     && $\mrm{~~~\vee\,\E{E}x(x\neq e1\wedge x\neq e2\,\wedge\,((s(x)=1\wedge none\neq none)\vee\,(s(x)=2\wedge none\neq none)}$\\
     && $~~~~~~~~~~~~~~~~~~~~~~~~~~~~~~~~~~~~~~~~\mrm{\vee\,(s(x)=3\wedge none\neq none)}$\\
     && $~~~~~~~~~~~~~~~~~~~~~~~~~~~~~~~~~~~~~~~~\mrm{\vee\,(s(x)\neq 1\wedge s(x)\neq 2\wedge s(x)\neq 3\wedge m_V(s(x))\neq none))))}$\\
    
     & $\equiv$ & $\mrm{\neg\E{E}x(x\neq e1\wedge x\neq e2\wedge s(x)\neq 1\wedge s(x)\neq 2\wedge s(x)\neq 3\wedge m_V(s(x))\neq none)}$
    \end{tabular}\\
    Here, we replace the terms $\mrm{s(e1),s(e2)}$ with node constant $\mrm{1}$, then replace $\mrm{m_V(1), m_V(2), m_V(3)}$ with $\mrm{none}$. Then, we simplify the resulting condition by evaluating $\mrm{none\neq none}$ which is equivalent to $\mrm{false}$.

\item 
    Val$(\Gamma_1, \mtt{del})$ = $\mrm{d\geq e}$ (for this case, we change nothing.)
\end{enumerate}
\end{example}

Finally, we define the transformation Lift, which takes a precondition and a generalised rule as an input and gives a left-application condition as an output. The output should express the precondition, the dangling condition, and the left-application condition that is given by the generalised rule.

\Def{Transformation Lift}{def:lift}{
For a precondition $c$ and a generalised rule $w=\tuple{r,ac_L,ac_R}$ with an unrestricted rule schema $r=\tuple{L\leftarrow K\rightarrow R}$,\\ $~~~~~~~~~~~~~~~~~~~~~~~~~~~~~~~~~~~~~\text{Lift}(c, w)=\text{Val}(\text{Split}(c\wedge ac_L, r)\wedge \text{Dang}(r), r).$}

\begin{example}[Transformation Lift]\normalfont
$~$\\
Lift$(q, \mtt{del}^\vee) = \mrm{\neg\E{E}x(x\neq e1\wedge x\neq e2\wedge s(x)\neq 1\wedge s(x)\neq 2\wedge s(x)\neq 3\wedge m_V(s(x))\neq none)}\wedge\,\mrm{d\geq e}$
\end{example}

\subsection{From Left to Right-Application Condition}

To obtain a right-application condition from the obtained left-application condition, we need to consider properties that could be different in the initial and result graphs. Recall that in constructing a left-application condition, we evaluate all functions with a node/edge constant argument so that the satisfaction of the condition is no longer independent of the properties of the left-hand graph.

The Boolean value for $\mrm{x=i}$ for any node/edge variable $x$ and node/edge constant $i$ not in $R$ must be false in the resulting graph. Analogously, $\mrm{x\neq i}$ is always true. Also, all variables in the left-application condition should not represent any new node and edge in the right-hand side. Hence, to obtain the right-application condition Shift$(c,w)$, we have some adjustment to the obtained left-application condition, denoted by Adj$(d,r)$ where $d=\text{Lift}(c,w)$.

To obtain Adj$(d,r)$, we follow the following steps:
\vspace{-\topsep}\begin{enumerate}
\item Replace every term representing indegree or outdegree if any (see \cite{WP20a} for detail);
\item Replace every subformula in the form $x_1\neq x_2$ with $\mrm{true}$ and $x_1= x_2$ with $\mrm{false}$ if $x_1$ or $x_2$ is in $V_L-V_K$ or $E_L-E_K$;
\item Replace every $\E{V}\mrm{x}(c_1)$ with $\E{V}\mrm{x}(x\neq v_1\wedge\ldots\wedge x\neq v_n\wedge c_1)$ and every $\E{E}\mrm{x}(c_1)$ with $\E{E}\mrm{x}(x\neq e_1$ $\wedge\ldots\wedge x\neq e_m\wedge c_1)$ for $V_R-V_K=\{v_1,\ldots,v_n\}$ and $E_R-E_K=\{e_1,\ldots,e_n\}$.
\end{enumerate}

\begin{definition}[Adjusment]\label{def:adj}\normalfont
Given an unrestricted rule schema $r=\tuple{L\leftarrow K\rightarrow R}$ and a condition $c$ over $L$. Let $c'$ be a condition over $L$ that is obtained from $c$ by changing every term $\mrm{incon(x)}$ (or $\mrm{outcon(x)}$) for $x\in V_K$ with $\mrm{indeg(x)-}indeg_R(x)$ (or $\mrm{outdeg(x)-}outdeg_R(x)$). Let also $\{v_1,\ldots,v_n\}$ and $\{e_1,\ldots,e_m\}$ denote the set of all nodes and edges in $R-K$ respectively. 
The \textit{adjusted} condition of $c$ w.r.t $r$, denoted by Adj$(c, r)$,  is a condition over $R$ that is defined inductively, where $c_1,c_2$ are conditions over $L$:
\begin{small}
\vspace{-\topsep}\begin{enumerate}\setlength{\parskip}{0pt} \setlength{\itemsep}{0pt plus 1pt}
    \item If $c$ is $\mrm{true}$ or $\mrm{false}$, Adj$(c,r)=c'$;
    \item If $c$ is the predicates $\mrm{int(x),char(x),string(x)}$ or $\mrm{atom(x)}$ for a list variable $x$, Adj$(c,r)=c'$;
    \item If $c=\mrm{root(x)}$ for some term $x$ representing a node,
    Adj$(c,r)=c'$
    \item If $c=x_1\ominus x_2$ for some terms $x_1, x_2$ and $\ominus\in\{=,\neq,<,\leq,>,\geq\}$,\\
    Adj$(c,r)$ =
            $\begin{cases} 
      \mrm{false} & ,\text{if $\ominus\in\{=\}$ and $x_1\in V_L-V_K\cup E_{L}$ or $x_2\in V_L-V_K\cup E_{L}$}, \\
        \mrm{true} & ,\text{if $\ominus\in\{\neq\}$ and $x_1\in V_L-V_K\cup E_{L}$ or $x_2\in V_L-V_K\cup E_{L}$}, \\
      c' & ,\text{otherwise}
   \end{cases}$
       \item Adj$(c_1\vee c_2,r) = \text{Adj}({c_1},r)\vee\text{Adj}({c_2},r)$
    \item  Adj$(c_1\wedge c_2,r) = \text{Adj}({c_1},r)\wedge\text{Adj}({c_2},r)$
    \item Adj$(\neg c_1,r)=\neg\text{Adj}({c_1},r)$
    \item Adj$(\E{V}\mrm{x}(c_1),r)=\E{V}\mrm{x}(x\neq v_1\wedge\ldots\wedge x\neq v_n\wedge\text{Adj}(c_1,r))$
    \item Adj$(\E{E}\mrm{x}(c_1),r)=\E{E}\mrm{x}(x\neq e_1\wedge\ldots\wedge x\neq e_m\wedge\text{Adj}(c_1,r))$
    \item Adj$(\E{L}\mrm{x}(c_1),r)=\E{L}\mrm{x}(\text{Adj}({c_1},r))$
   \hfill$\square$
\end{enumerate}\vspace{-\topsep}\end{small}
\end{definition}

\begin{example}[Adjusment]\normalfont$~$\\
Let $p$ denotes Lift$(q,\mtt{del^\vee})$. Then,\\ 
Adj$(p,\mtt{del})$ = $\mrm{\neg\E{E}x(x\neq e1\wedge s(x)\neq 1\wedge s(x)\neq 2\wedge m_V(s(x))\neq none)\wedge\,\mrm{d\geq e}}$
\end{example}

Although Adj$(\text{Lift}(c,w),r)$ can be considered as a right-application condition, we need a stronger condition to have a strongest liberal postcondition. Hence, we add a condition over $R$ expressing the specification of the right-hand graph. A \textit{specification of a graph} $R$, denoted by Spec$(R)$, can be easily obtained by forming conjunction of predicates, equality of functions and their value in $R$, and type of label variables in $R$.

\Def{Specifying a Totally Labelled Graph}{def:spec}{
Given a totally labelled graph $R$ with the set of nodes $V_R=\{v_1,\ldots,v_n\}$ and the set of edges $E_R=\{e_1,\ldots,e_m\}$. Let $X=\{x_1,\ldots,x_k\}$ be the set of all list variables in $R$, and Type$(x)$ for $x\in X$ is $\mrm{int(x)}$, $\mrm{char(x)}$, $\mrm{string(x)}$, $\mrm{atom(x)}$, or $\mrm{true}$ if $x$ is an integer, char, string, atom, or list variable respectively. Let also Root$_R(v)$ for $v\in V_R$ be a function such that Root$_R(v)=\mrm{root(v)}$ if $p_R(v)=1$, and Root$_R(v)=\mrm{\neg root(v)}$ otherwise. \emph{A specification of $R$}, denoted by Spec$(R)$, is the condition over $R$:\\
$~~~~~~~~~~~~~~~~~~~~~~~~~\mrm{\bigwedge_{i=1}^k} \text{Type}(x_i) \mrm{~\wedge~\bigwedge_{i=1}^n \lV(v_i)}=\lst_R(v_i)\mrm{~\wedge~\mV(v_i)=}\mrk_R(v_i)\mrm{~\wedge~} \text{Root}_R(v_i)$\\
$~~~~~~~~~~~~~~~~~~~~~~~~~\mrm{\wedge~\bigwedge_{i=1}^m s(e_i)=}s_L(e_i)\mrm{~\wedge~ t(e_i)=}t_R(e_i)\mrm{~\wedge~\lE(e_i)}=\lst_L(e_i)\mrm{~\wedge~
\mE(e_i)=}\mrk_R(e_i)$
}

Basically, Spec$(R)$ explicitly shows us node and edge identifiers in $R$, label, mark, and rootedness of each node in $R$ (if defined), also the source, target, label, and mark of each edge in $R$.

\Lemma{lemma:spec}{
For every totally labelled rule graph $R$, there exists a condition Spec$(R)$ such that for every host graph $G$, $G\Sat$Spec$(R)$ if and only if there exists assignment $\alpha_\mathbb{L}$ such that $g:R^{\alpha_\mathbb{L}}\rightarrow G$ is an inclusion.
}

\Def{Shifting}{def:shift}{
Given a generalised rule $w=\tuple{r,ac_L,ac_R}$ for an unrestricted rule schema $r=\tuple{L\leftarrow K\rightarrow R}$, and a precondition $c$. Right application condition w.r.t. $c$ and $w$, denoted by Shift$(c,w)$, is defined as:\\
$~~~~~~~~~~~~~~~~~~~~~~~~~~~~~~~~~~~~~\text{Shift$(c,w)=$Adj(Lift$(c,w),r)\wedge\, ac_R\, \wedge\,$Spec$(R)\,\wedge\,$Dang$(r^{-1})$}.$
}

\begin{example}[Obtaining Right-Application Condition]\normalfont $~$\\
\begin{tabular}{lcp{10.5cm}}
    Shift$(q,\mtt{del}^\vee)$&=&$\mrm{\neg\E{E}x(x\neq e1\wedge s(x)\neq 1\wedge s(x)\neq 2\wedge m_V(s(x))\neq none)\wedge\,\mrm{d\geq e}}$\\
    &&$\mrm{\wedge\, \lV(1)=a\wedge\lV(2)=b\wedge\lE(e1)=d+e\wedge\mV(1)=red}$\\
    &&$\mrm{\wedge\mV(2)=none\wedge\mE(e1)=none\wedge s(e1)=1\wedge t(e1)=2}$\\
    &&$\mrm{\wedge \neg root(1)\wedge\neg root(2)\wedge int(d)\wedge int(e)}$
\end{tabular}
\end{example}

\subsection{From Right-Application Condition to Postcondition}

The right-application condition we obtained from transformation Shift is strong enough to express properties of the replacement graph of any resulting graph. To be able to check the satisfaction of the condition in the resulting graph, we need to change it to a FO formula. This can be done by replacing every node and edge constant to a fresh variable and state that each new variable is not equal to other new variables.

\Lemma{lemma:var}{
For a rule graph $G$ and a condition $c$ over $G$, there exists a first-order formula Var$(c)$ so that for every graph $H$ that is isomorphic to $G$, $G\Sat c$ implies $H\Sat$Var$(c)$.}

To obtain a closed FO formula from the obtained right-application condition, we only need to variablise the node/edge constants in the right-application condition, then put an existential quantifier for each free variable in the resulting FO formula. In \cite{WP20a}, we show that the obtained formula defines a strongest liberal postcondition.

\Def{Formula Post}{def:post}{
Given a generalised rule $w=\tuple{r,ac_L,ac_R}$ for an unrestricted rule $r=\tuple{L\leftarrow K\rightarrow R}$ and a precondition $c$. Let $\{x_1,\ldots,x_n\}$, $\{y_1,\ldots,y_m\}$, and $\{z_1,\ldots,z_k\}$ denote the set of free node, edge, and label (resp.) variables in Var(Shift$(c,w)$). We define Post$(c,w)$ as the FO formula:\\
$~~~~~~~~~~~~~~~~~~~~~~~~~~~~~~~~~~~~~\text{Post}(c, w)\equiv\E{V}x_1,\ldots,x_n(\E{E}y_1,\ldots,y_m(\E{L}z_1,\ldots,z_k(\text{Var(Shift}(c,w))))).$\\
For a rule schema $r$, we denote by Slp$(c,r)$ and Slp$(c,r^{-1})$ the formulas Post$(c,r^\vee)$ and Post$(c,(r^\vee)^{-1})$ respectively.}

\begin{example}[Obtaining Strongest Liberal Postcondition]\normalfont$~$\\
\begin{tabular}{lcp{10.5cm}}
    Slp$(q,\mtt{del})$&=&$\mrm{\E{V}u,v(u\neq v\wedge \E{E}w(\E{L}a,b,d,e(}$\\
    &&$\mrm{\neg\E{E}x(x\neq w\wedge s(x)\neq u\wedge s(x)\neq v\wedge m_V(s(w))\neq none)\wedge\,\mrm{d\geq e}}$\\
    &&$\mrm{\wedge\, \lV(u)=a\wedge\lV(v)=b\wedge\lE(w)=d+e\wedge\mV(u)=red}$\\
    &&$\mrm{\wedge\mV(v)=none\wedge\mE(w)=none\wedge s(w)=u\wedge t(w)=v}$\\
    &&$\mrm{\wedge \neg root(u)\wedge\neg root(v)\wedge int(d)\wedge int(e))))}$
\end{tabular}
\end{example}

\Theo{Strongest liberal postconditions}{theo:slp}{
Given a precondition $c$ and a conditional rule schema $r=\tuple{\langle L \leftarrow K\rightarrow R\rangle,\Gamma}$. Then, Slp$(c,r)$ is a strongest liberal postcondition w.r.t. $c$ and $r$.}

\section{Proof Calculi}
\label{sec:proofrules}
In this section, we introduce a semantic and a syntactic partial correctness calculus. As pre- and postconditions, we use arbitrary assertions for the former, and first-order formulas for the latter. 

Given a graph program $P$ and assertions $c$ and $d$, a triple $\{c\}\,P\,\{d\}$ is \emph{partially correct}, denoted by $\vDash \{c\}~P~\{d\}$, if for every graph $G$ satisfying $c$, all graphs in $\Sem{P}G$ satisfy $d$ \cite{Poskitt-Plump10a}.

\subsection{Semantic Partial Correctness Calculus}
\label{sec:sem}

Besides strongest liberal postconditions, it will be useful to consider weakest liberal preconditions.

\Def{Weakest liberal precondition}{def:wlpP}{
An assertion $c$ is a \emph{liberal precondition} with respect to a graph program $P$ and a postcondition $d$, if for all host graphs $G$ and $H$,\\ 
$~~~~~~~~~~~~~~~~~~~~~~~~~~~~~~~~~~~~~~~~~~~~~~~~~~~~~~~G\vDash c \text{ and } H\in\Sem{P}G \text{ implies }H\vDash d.$\\
A \emph{weakest liberal precondition} w.r.t. $P$ and $d$, written WLP$(P,d)$, is a liberal precondition w.r.t. $P$ and $d$ that is implied by all liberal postconditions w.r.t. $P$ and $d $.}

To prove that a triple $\{c\}~P~\{d\}$ is partially correct, we only need to show that SLP$(c,P)$ implies $d$ or WLP($P,d$) implies $c$. However, if $P$ contains a loop, obtaining SLP$(c,P)$ or WLP$(P,d)$ may be difficult because $P$ may diverge. In \cite{HP09,Pennemann09}, divergence is represented by infinite formulas while in \cite{Jones-Roscoe-Wood10a} approximations of these assertions are used. We take a different approach by considering SLP and WLP only for loop-free programs. Programs with loops are verified using the proof rule [alap] in the calculi introduced below.

Before we define our proof rules, we define assertions expressing that a program can produce a result graph or may fail, respectively. These assertions are needed in the proof rules for the branching commands $\mtt{if\_then\_else}$ and $\mtt{try\_then\_else}$.

\Def{Assertions SUCCESS and FAIL}{def:assSEFE}{
For a graph program $P$, SUCCESS$(P)$ and FAIL$(P)$ are the predicates defined on all host graphs $G$ by\\ $G\Sat\text{SUCCESS}(P)\text{~if and only if there exists a host graph $H$ with~} H\in\llbracket P\rrbracket G,$ and
\\$G\Sat\text{FAIL}(P)\text{~if and only if~} \text{fail}\in\llbracket P\rrbracket G.$}

We also define a predicate Break to deal with loops containing the $\mtt{break}$ command. 
\Def{Predicate Break}{def:Break}{
Given a graph program $P$ and assertions $c$ and $d$, Break$(c,P,d)$  holds if and only if for all derivations $\langle P, G\rangle \rightarrow^* \langle\mtt{break},H\rangle$, $G\Sat c$ implies $H\Sat d$.}

Here $P$ is a loop body whose execution on graph $G$ encounters the $\mtt{break}$ command, and $H$ is the graph that has been reached at that point. 

\Def{Semantic partial correctness proof rules}{def:semproofrule}{
The semantic partial correctness proof rules for GP\,2 commands, denoted by \textrm{SEM}, are defined in \figurename~\ref{fig:semprules}, where $c, d,$ and $d'$ are assertions, $r$ is a conditional rule schema, $\mathcal{R}$ is a set of rule schemata, and $C, P$, and $Q$ are graph programs.}

The assertions SUCCESS and FAIL are needed to prove a triple about an $\mtt{if}$ command, because $P$ may be executed on $G$ if $G\Sat\SUCCESS(C)$, and $Q$ may be executed on $G$ if $G\Sat\FAIL(C)$. Similarly, for a $\mtt{try}$ command, $P$ may be executed on a graph $C'$ if $G\Sat\SUCCESS(C)$ and $C'\in\Sem{C}G$, and $Q$ may be executed on $G$ if $G\Sat\FAIL(C)$. Finally the execution of a loop $P!$, it terminates if at some point the execution of $P$ yields failure, or reaches the command $\mtt{break}$.\\

\begin{figure}[htb]
\begin{subfigure}{.5\textwidth}
\centering
\footnotesize
\def\arraystretch{3}\tabcolsep=2pt
~[ruleapp]$_{\text{slp}}~\displaystyle\frac{}{\{c\}~r~\{\text{SLP}(c,r)\}}$\\$~$\\
~[ruleapp]$_{\text{wlp}}~\displaystyle\frac{}{\{\text{WLP}(r,d)\}~r~\{d\}}$\\$~$\\
~[ruleset]$~\displaystyle\frac{\{c\}~r~\{d\}\text{ for each }r\in\mathcal{R}}{\{c\}~\mathcal{R}~\{d\}}$\\$~$\\
~[comp]$\displaystyle\frac{\{c\}~P~\{e\}~~~~\{e\}~P~\{d\}}{\{c\}~P;Q~\{d\}}$
\\$~$\\
~[cons]~$\displaystyle\frac{c\text{ implies }c'~~~\{c'\}~P~\{d'\}~~~d'\text{ implies }d}{\{c\}~P~\{d\}}$\\$~$\\
{~[if]$~\displaystyle\frac{\{c\wedge\text{S}(C)\}~P~\{d\}~~~\{c\wedge\text{F}(C)\}~Q~\{d\}}{\{c\}~\mtt{if~}C\mtt{~then~}P\mtt{~else~}Q~\{d\}}$}\\$~$\\
{~[try]$~\displaystyle\frac{\{c\wedge\text{S}(C)\}~C;P~\{d\}~~~\{c\wedge\text{F}(C)\}~Q~\{d\}}{\{c\}~\mtt{try~}C\mtt{~then~}P\mtt{~else~}Q~\{d\}}$}
\\$~$\\
~[alap]$~\displaystyle\frac{\{c\}~P~\{c\}~~~~~\text{Break}(c,P,d)}{\{c\}~P!~\{(c\wedge\text{F}(P))\vee d\}}$
 \caption{Calculus {SEM}}
   \label{fig:semprules}
\end{subfigure}
\begin{subfigure}{.5\textwidth}
  \centering
      \footnotesize
\def\arraystretch{3}\tabcolsep=2pt
~[ruleapp]$_{\text{slp}}~\displaystyle\frac{}{\{c\}~r~\{\text{Slp}(c,r)\}}$\\$~$\\
~[ruleapp]$_{\text{wlp}}~\displaystyle\frac{}{\{\neg \text{Slp}(\neg d,r^{-1})\}~r~\{d\}}$\\$~$\\
~[ruleset]$~\displaystyle\frac{\{c\}~r~\{d\}\text{ for each }r\in\mathcal{R}}{\{c\}~\mathcal{R}~\{d\}}$\\$~$\\
~[comp]$\displaystyle\frac{\{c\}~P~\{e\}~~~~\{e\}~P~\{d\}}{\{c\}~P;Q~\{d\}}$
\\$~$\\
~[cons]~$\displaystyle\frac{c\text{ implies }c'~~~\{c'\}~P~\{d'\}~~~d'\text{ implies }d}{\{c\}~P~\{d\}}$\\$~$\\
{~[if]$~\displaystyle\frac{\{c\wedge\text{Success}(C)\}~P~\{d\}~~~\{c\wedge\text{Fail}(C)\}~Q~\{d\}}{\{c\}~\mtt{if~}C\mtt{~then~}P\mtt{~else~}Q~\{d\}}$}\\$~$\\
{~[try]$~\displaystyle\frac{\{c\wedge\text{Success}(C)\}~C;P~\{d\}~~~\{c\wedge\text{Fail}(C)\}~Q~\{d\}}{\{c\}~\mtt{try~}C\mtt{~then~}P\mtt{~else~}Q~\{d\}}$}
\\$~$\\
~[alap]$~\displaystyle\frac{\{c\}~S~\{c\}~~~~~~~~~\text{Break}(c,S,d)}{\{c\}~S!~\{(c\wedge\text{Fail}(S))\vee d\}}$
    \caption{Calculus {SYN}}
    \label{fig:synprules}
\end{subfigure}
\caption{Semantic (a) and syntactic (b) partial correctness proof calculus, where S$(C)$ is SUCCESS($C$) and F$(C)$ is FAIL$(C)$}
\label{fig:fig}
\end{figure}

\subsection{Syntactic Partial Correctness Calculus}




Defining a first-order formula for SUCCESS$(r)$ with a rule schema $r$ is easier than defining FO formula for SUCCESS$(P)$ with a program $P$ with loops. This is because the existence of a result graph can be known after some execution of $P$, which really depends on the program. Moreover, it may diverge. However if we consider loop-free programs, we can construct a first-order formula for SUCCESS, FAIL and SLP. In addition, we can construct a FO formula of FAIL($P$) for bigger class of programs because some commands cannot fail (see \cite{Bak15a}).

\begin{definition}[Non-failing commands]\label{def:nofail}\normalfont
The class of \emph{non-failing commands} is inductively defined as follows:
\vspace{-\topsep}\begin{enumerate}\setlength{\parskip}{0pt} \setlength{\itemsep}{0pt plus 1pt}
    \item $\mtt{break}$ and $\mtt{skip}$ are non-failing commands
    \item Every call of a rule schema with the empty graph as its left-hand graph is a non-failing command
    \item Every rule set call $\{r_1,\ldots,r_n\}$ for $n\geq 1$ where each $r_i$ has the empty graph as its left-hand graph, is a non-failing command
    \item Every command P! is a non-failing command
    \item if $P$ and $Q$ are non-failing commands, then $P;Q$, $\mtt{if\,}C\mtt{\,then\,}P\mtt{\,else\,}Q$, and $\mtt{try\,}C\mtt{\,then\,}P\mtt{\,else\,}Q$ are non-failing commands.\hfill$\square$
\end{enumerate}
\end{definition}

Now, let us consider $P$ in the form $C;Q$. For any host graph $G$, $\text{fail}\in\Sem{C;Q}G$ iff $\text{fail}\in\Sem{C}G$ or $H\in\Sem{C}G\wedge\text{fail}\in\Sem{Q}H$ for some host graph $H$, which means $G\Sat \FAIL(C)\vee(\SUCCESS(C)\wedge\FAIL(Q))$. We can construct both Fail$(C)$ and Success$(C)$ if $C$ is a loop-free program (see \cite{WP20a} for the detail of construction), and we can construct Fail$(Q)$ if $Q$ is a loop-free program or a non-failing command. Here, we introduce the class of \textit{iteration commands} for which we can obtain Fail of the commands.

\begin{definition}[Iteration commands]\label{def:iteration}\normalfont
The class of iteration commands is inductively defined as follows:
1) every loop-free program and non-failing command is an iteration command, and 2) a command in the form $C;P$ is an iteration command if $C$ is a loop-free program and $P$ is an iteration command.\hfill$\square$
\end{definition}

If $S$ is a loop-free program, we can construct Fail$(S)$ as stated above (see the full construction in \cite{WP20a}. Meanwhile, if $S$ is a non-failing command, there is no graph $G$ such that fail$\in\Sem{S}G$, so we can conclude that Fail$(S)\equiv\mrm{false}$. If $S$ is in the form of $C;P$ for a loop-free program $C$ and a non-failing program $P$, fail$\in\Sem{S}G$ for a graph $G$ only if fail$\in\Sem{C}G$ (because $P$ cannot fail), so that Fail$(S)\equiv$ Fail$(C)$.

\begin{definition}\label{def:Failit}\normalfont
Let Fail$_{\text{lf}}(C)$ denotes the formula Fail$(C)$ for a loop-free program $C$. For any iteration command $S$,\\
\footnotesize{Fail$(S)=\begin{cases}
\mrm{false} & \text{if $S$ is a non-failing command}\\
\text{Fail}_{\text{lf}}(S) & \text{if $S$ is a loop-free program}\\
\text{Fail}(C) & {\text{if $S=C;P$ for a loop-free program $C$, a non-failing program $P$}}\\
\end{cases}$\hfill$\square$}
\end{definition}

\begin{theorem}\normalfont\label{theo:nofail}
For any loop-free program $P$ and precondition $c$, there exists first-order formula Success$(P)$ and Slp$(c,P)$ such that $G\Sat\text{Success}(P)$ if and only if $G\Sat\text{SUCCESS}(P)$ and $G\Sat\text{Slp}(c,P)$ if and only if $G\Sat\text{SLP}(c,P)$.
Also, for any iteration command $S$, $\text{$G\Sat\text{Fail}(S)$ if and only if $G\Sat\text{FAIL}(S)$}$.
\end{theorem}

The construction of Slp($c,P)$ and Success$(P)$ to show that Theorem \ref{theo:nofail} holds can be found in \cite{WP20a}. Since we only have a construction for Success$(C)$ for a loop-free program $C$ and Fail$(S)$ for an iteration command $S$, we cannot define the syntactic proof calculus for arbitrary graph programs. We call the class of programs we can handle by our syntactical calculus as control programs. 

\Def{Control programs}{def:controlprogram}{
A \emph{control command} is a command where the condition of every branching command (e.g. the command $C$ of \ttt{if} $C$ \ttt{then} $P$ \ttt{else} $Q$) is loop-free and every loop body is an iteration command. Similarly, a graph program is a \emph{control program} if all its command are control commands.}

As in \cite{HP09}, a First-order formula of WLP($r,d$) of a postcondition $d$ and a rule schema $r$ can be easily constructed from the construction of a strongest liberal postcondition.

\begin{lemma}\label{lemma:wlpr}\normalfont
Given a closed FO formula $d$ and a rule schema $r$. Then for all host graphs $G$,\\
$G\Sat\neg\text{Slp}(\neg d,r^{-1}) \text{ if and only if } G\Sat\text{WLP}(r,d).$
\end{lemma}

\Def{Syntactic partial correctness proof rules}{def:synproofrule}{
The syntactic partial correctness proof rules, denoted by {SYN}, are defined in \figurename~\ref{fig:synprules}, where $c, d,$ and $d'$ are conditions, $r$ is a conditional rule schema, $\mathcal{R}$ is a set of rule schemata, $C$ is a loop-free program, $P$ and $Q$ are control commands, and $S$ is an iteration command.}

In the following section, we give a graph verification example using the calculus {SYN} we defined in this section.

\section{Example: Verifying a 2-Colouring Program}
\label{sec:ex}
\begin{figure}
\begin{scriptsize}\begin{tabular}{l}
\scriptsize{$\mtt{Main = (init; Colour!)!; if~Illegal~then~unmark!}$}\\
$\mtt{Colour = \{col\_\,blue, col\_\,red\}}$\\
$\mtt{Illegal = \{ill\_\,blue, ill\_\,red\}}$\\
\end{tabular}\\[1.5ex]
\begin{tabular}{lllll}
\begin{tikzpicture}[remember picture,
  inner/.style={circle,draw,minimum size=16pt},
  outer/.style={inner sep=2pt}, scale=0.6
  ]
  \node[outer] (AA) at (0,1) {$\tiny\mtt{init(a:list)}$};
  \node[outer] (A) at (-1,0) {
   \begin{tikzpicture}[scale=0.8, transform shape]
		\node[inner, label=below:\tiny 1] (Aa) at (0,0) {$\mtt{a}$};
		\end{tikzpicture}};
  \node[outer] (B) at (0,0) {$\Rightarrow$};
  \node[outer] (C) at (1,0) {
   \begin{tikzpicture}[scale=0.8, transform shape]
		\node[inner, label=below:\tiny 1, fill=red!70] (Aa) at (0,0) {$\mtt{a}$};	
		\end{tikzpicture}};
	\end{tikzpicture}
	
	&$~~~~~~~~~$&\begin{tikzpicture}[remember picture,
  inner/.style={circle,draw,minimum size=16pt},
  outer/.style={inner sep=2pt}, scale=0.6
  ]
  \node[outer] (AA) at (0,1) {$\tiny\mtt{col\_\,blue(a,b,c:list)}$};
  \node[outer] (A) at (-1,0) {
   \begin{tikzpicture}[scale=0.8, transform shape]
		\node[inner, label=below:\tiny 1, fill=red!70] (Aa) at (0,0) {$\mtt{a}$};
		\node[inner, label=below:\tiny 2] (Ab) at (1,0) {$\mtt{b}$};
		\draw (Aa) to node[above] {$\mtt{c}$} (Ab);
		\end{tikzpicture}};
  \node[outer] (B) at (0.5,0) {$\Rightarrow$};
  \node[outer] (C) at (2,0) {
   \begin{tikzpicture}[scale=0.8, transform shape]
		\node[inner, label=below:\tiny 1, fill=red!70] (Aa) at (0,0) {$\mtt{a}$};
		\node[inner, label=below:\tiny 2, fill=blue!50] (Ab) at (1,0) {$\mtt{b}$};
		\draw (Aa) to node[above] {$\mtt{c}$} (Ab);	
		\end{tikzpicture}};
	\end{tikzpicture}
&$~~~~~~~$&
\begin{tikzpicture}[remember picture,
  inner/.style={circle,draw,minimum size=16pt},
  outer/.style={inner sep=2pt}, scale=0.6
  ]
  \node[outer] (AA) at (0,1) {$\tiny\mtt{col\_\,red(a,b,c:list)}$};
  \node[outer] (A) at (-1,0) {
   \begin{tikzpicture}[scale=0.8, transform shape]
		\node[inner, label=below:\tiny 1, fill=blue!50] (Aa) at (0,0) {$\mtt{a}$};
		\node[inner, label=below:\tiny 2] (Ab) at (1,0) {$\mtt{b}$};
		\draw (Aa) to node[above] {$\mtt{c}$} (Ab);
		\end{tikzpicture}};
  \node[outer] (B) at (0.5,0) {$\Rightarrow$};
  \node[outer] (C) at (2,0) {
   \begin{tikzpicture}[scale=0.8, transform shape]
		\node[inner, label=below:\tiny 1, fill=blue!50] (Aa) at (0,0) {$\mtt{a}$};
		\node[inner, label=below:\tiny 2, fill=red!70] (Ab) at (1,0) {$\mtt{b}$};
		\draw (Aa) to node[above] {$\mtt{c}$} (Ab);	
		\end{tikzpicture}};
	\end{tikzpicture}\\

\begin{tikzpicture}[remember picture,
  inner/.style={circle,draw,minimum size=16pt},
  outer/.style={inner sep=2pt}, scale=0.6
  ]
  \node[outer] (AA) at (0,1) {$\tiny\mtt{unmark(a:list)}$};
  \node[outer] (A) at (-1,0) {
   \begin{tikzpicture}[scale=0.8, transform shape]
		\node[inner, label=below:\tiny 1, fill=magenta!70] (Aa) at (0,0) {$\mtt{a}$};
		\end{tikzpicture}};
  \node[outer] (B) at (0,0) {$\Rightarrow$};
  \node[outer] (C) at (1,0) {
   \begin{tikzpicture}[scale=0.8, transform shape]
		\node[inner, label=below:\tiny 1] (Aa) at (0,0) {$\mtt{a}$};
		\end{tikzpicture}};
	\end{tikzpicture}
	&&
\begin{tikzpicture}[remember picture,
  inner/.style={circle,draw,minimum size=16pt},
  outer/.style={inner sep=2pt}, scale=0.6
  ]
  \node[outer] (AA) at (0,1) {$\tiny\mtt{ill\_\,blue(a,b,c:list)}$};
  \node[outer] (A) at (-1,0) {
   \begin{tikzpicture}[scale=0.8, transform shape]
		\node[inner, label=below:\tiny 1, fill=blue!50] (Aa) at (0,0) {$\mtt{a}$};
		\node[inner, label=below:\tiny 2, fill=blue!50] (Ab) at (1,0) {$\mtt{b}$};
		\draw (Aa) to node[above] {$\mtt{c}$} (Ab);
		\end{tikzpicture}};
  \node[outer] (B) at (0.5,0) {$\Rightarrow$};
  \node[outer] (C) at (2,0) {
   \begin{tikzpicture}[scale=0.8, transform shape]
		\node[inner, label=below:\tiny 1, fill=blue!50] (Aa) at (0,0) {$\mtt{a}$};
		\node[inner, label=below:\tiny 2, fill=blue!50] (Ab) at (1,0) {$\mtt{b}$};
		\draw (Aa) to node[above] {$\mtt{c}$} (Ab);	
		\end{tikzpicture}};
	\end{tikzpicture}
&&
\begin{tikzpicture}[remember picture,
  inner/.style={circle,draw,minimum size=16pt},
  outer/.style={inner sep=2pt}, scale=0.6
  ]
  \node[outer] (AA) at (0,1) {$\tiny\mtt{ill\_\,red(a,b,c:list)}$};
  \node[outer] (A) at (-1,0) {
   \begin{tikzpicture}[scale=0.8, transform shape]
		\node[inner, label=below:\tiny 1, fill=red!70] (Aa) at (0,0) {$\mtt{a}$};
		\node[inner, label=below:\tiny 2, fill=red!70] (Ab) at (1,0) {$\mtt{b}$};
		\draw (Aa) to node[above] {$\mtt{c}$} (Ab);
		\end{tikzpicture}};
  \node[outer] (B) at (0.5,0) {$\Rightarrow$};
  \node[outer] (C) at (2,0) {
   \begin{tikzpicture}[scale=0.8, transform shape]
		\node[inner, label=below:\tiny 1, fill=red!70] (Aa) at (0,0) {$\mtt{a}$};
		\node[inner, label=below:\tiny 2, fill=red!70] (Ab) at (1,0) {$\mtt{b}$};
		\draw (Aa) to node[above] {$\mtt{c}$} (Ab);	
		\end{tikzpicture}};
	\end{tikzpicture}
	
\end{tabular}\end{scriptsize}
    \caption{Graph program \ttt{2-colouring}}
    \label{fig:2col}
\end{figure}
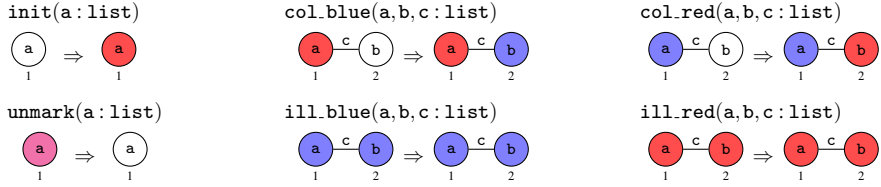

In this section, we show how to verify the 2-colouring graph program given in \figurename~\ref{fig:2col}. The 2-colouring problem is the problem to assign to each node of a graph one of two colours such that each two adjacent nodes have different colours. 

The program expects input graphs without any roots or marks. It starts by marking any unmarked node with red, then repeatedly colours uncoloured nodes adjacent to a coloured node  with the other colour. Finally, the program checks if the produced graph contains two adjacent nodes with the same colour. If that is the case, the program unmarks all nodes to restore the input graph. Note the nested loop which allows to process disconnected graphs, by colouring each connected component in turn. This program cannot be verified with the proof calculi in \cite{PoskittP12,Poskitt13} as there exists a nested loop in the program.

Let us consider the precondition ``every node and edge is unmarked and every node is unrooted" and the postcondition ``the precondition holds or every node is marked with blue or red, and no two adjacent nodes marked with the same colour", that can be represented by $c$ and $c\vee d$ where\\
\begin{footnotesize}$c=\mrm{\A{V}x(\mV(x)= none\wedge\neg root(x))\wedge\A{E}x(\mE(x)=none)}$, and\\
$d=\mrm{\A{V}x((\mV(x)=red\vee\mV(x)=blue))\wedge\neg\E{E}x(s(x)\neq t(x)\wedge\mV(s(x))=\mV(t(x)))}$\end{footnotesize}

By using the conditions in Table \ref{tab:2col-cond}, we then have a proof tree as in \figurename~\ref{fig:2col-proof} for the partial correctness of $\mtt{2-colouring}$ with respect to $c$ and $c\vee d$.

\begin{table}[]
    \caption{Conditions inside proof tree of $\mtt{2-colouring}$}
    \label{tab:2col-cond}
    \centering\begin{scriptsize}
    \begin{tabular}{|p{15.5cm}|}
        \hline
        \multicolumn{1}{|c|}{\textbf{symbol and its first-order formulas}}\\\hline
        
        $c
        \equiv
        \mrm{\A{V}x(\mV(x)=none\wedge\neg root(x))\wedge\A{E}x(\mE(x)=none)}$\\\hline
        
        $d\equiv
        \mrm{\A{V}x((\mV(x)=red\vee\mV(x)=blue))\wedge\neg\E{E}x(s(x)\neq t(x)\wedge\mV(s(x))=\mV(t(x)))}$\\\hline
        
        $e
        \equiv
        \mrm{\A{V}x((\mV(x)=red\vee\mV(x)=blue)\wedge\neg root(x))\wedge\A{E}x(\mE(x)=none)}$\\\hline
        
        $f
        \equiv
        \mrm{\A{V}x((\mV(x)=red\vee\mV(x)=blue\vee\mV(x)=none))\wedge\neg root(x))\wedge\A{E}x(\mE(x)=none)}$\\\hline
       
        Slp$(f,\mtt{init})$\\
        $\equiv
        \mrm{\E{V}y(\A{V}x(x=y\vee((\mV(x)=red\vee\mV(x)=blue\vee\mV(x)=none)\wedge\neg root(x)))\wedge \mV(y)=red\wedge\neg root(y))\wedge\A{E}x(\mE(x)=none)}$\\\hline
        
        Slp$(f,\mtt{c\_blue})=$Slp$(f,\mtt{c\_red})$\\
        $\equiv
        \mrm{\E{V}u,v(\A{V}x(x=u\vee x=v\vee((\mV(x)=red\vee\mV(x)=blue\vee\mV(x)=none)\wedge\neg root(x)))}$\\
       $\mrm{~~~~~~~~~\wedge \mV(u)=red\wedge\mV(v)=blue\wedge\neg root(u)\wedge\neg root(v)\wedge\E{E}y((s(y)=u\wedge t(y)=v)\vee(t(y)=u\wedge s(y)=v)))\wedge\A{E}x(\mE(x)=none)}$\\\hline
        
        Slp(f,\ttt{unmark})\\
        $\equiv
        \mrm{\E{V}y(\A{V}x(x=y\vee((\mV(x)=red\vee\mV(x)=blue\vee\mV(x)=none)\wedge\neg root(x)))\wedge\mV(y)=none\wedge\neg root(y))\wedge\A{E}x(\mE(x)=none)}$\\\hline
        
        Fail$(\mtt{Colour})$\\
        $\equiv
        \mrm{\neg\E{E}x((((\mV(s(x))=red\vee\mV(s(x))=blue)\wedge \mV(t(x))=none)\vee((\mV(t(x))=red\vee\mV(t(x))=blue)\wedge \mV(s(x))=none))}$\\
        $\mrm{~~~~~~~~~~~~~~~~~\wedge \neg root(s(x))\wedge\neg root(t(x))})$\\\hline
        
        Fail$(\mtt{init;Colour!})
        \equiv
        \mrm{\neg\E{V}x(\mV(x)=none\wedge\neg root(x))}$\\\hline

        Fail$(\mtt{unmark})
        \equiv
        \mrm{\neg\E{V}x(\mV(x)\neq none\wedge\neg root(x))}$\\\hline

        Fail$(\mtt{Illegal})\equiv
        \mrm{\neg\E{E}x(s(x)\neq t(x)\wedge((\mV(s(x))=red\wedge\mV(t(x))=red)\vee(\mV(s(x))=blue\wedge\mV(t(x))=blue)))}$\\\hline

        Success$(\mtt{Illegal})\equiv
        \mrm{\E{E}x(s(x)\neq t(x)\wedge((\mV(s(x))=red\wedge\mV(t(x))=red)\vee(\mV(s(x))=blue\wedge\mV(t(x))=blue)))}$\\\hline

    \end{tabular}\end{scriptsize}
\end{table}

\begin{figure}[h!]
 \begin{scriptsize}
    
\begin{prooftree}
\AxiomC{Subtree I}
\AxiomC{Subtree II}
\LeftLabel{[comp]}
\BinaryInfC{\{$f$\}$~\mtt{2colouring}~$\{$c\vee d$\}}
\LeftLabel{[cons]}
\UnaryInfC{\{$c$\}$~\mtt{2colouring}~$\{$c\vee d$\}}
\end{prooftree}

where subtree I is:
\begin{prooftree}
\AxiomC{}
\LeftLabel{[ruleapp]$_\text{slp}$}
\UnaryInfC{\{$f$\}$\mtt{init}$\{Slp($f,\mtt{init})$\}}
\LeftLabel{[cons]}
\UnaryInfC{\{$f$\}$\mtt{init}$\{$f$\}}

\AxiomC{}
\LeftLabel{[ruleapp]$_\text{slp}$}
\UnaryInfC{\{$f$\}$~\mtt{c\_\,blue}~$\{Slp$(f,\mtt{c\_\,blue})$\}}
\LeftLabel{[cons]}
\UnaryInfC{\{$f$\}$~\mtt{c\_\,blue}~$\{$f$\}}

\AxiomC{}
\LeftLabel{[ruleapp]$_\text{slp}$}
\UnaryInfC{\{$f$\}$~\mtt{c\_\,red}~$\{Slp$(f,\mtt{c\_\,red})$\}}
\LeftLabel{[cons]}
\UnaryInfC{\{$f$\}$~\mtt{c\_\,red}~$\{$f$\}}

\LeftLabel{[cons]}
\BinaryInfC{\{$f$\}$~\mtt{Colour}~$\{$f$\}}
\LeftLabel{[alap]}
\UnaryInfC{\{$f$\}$~\mtt{Colour!}~$\{$f\wedge\text{Fail}(\mtt{Colour})$\}}
\LeftLabel{[cons]}
\UnaryInfC{\{$f$\}$~\mtt{Colour!}~$\{$f$\}}
\LeftLabel{[comp]}
\BinaryInfC{\{$f$\}$~\mtt{init;Colour!}~$\{$f$\}}
\LeftLabel{[alap]}
\UnaryInfC{\{$f$\}$~\mtt{(init; Colour!)!~}$\{$f\wedge\text{Fail}(\mtt{init;Colour!})$\}}
\LeftLabel{[cons]}
\UnaryInfC{\{$f$\}$~\mtt{(init;Colour!)!}~$\{$e$\}}

\end{prooftree}

and subtree II is:
\begin{prooftree}
\AxiomC{}
\LeftLabel{[ruleapp]$_\text{slp}$}
\UnaryInfC{\{$f$\}$~\mtt{unmark}~$\{Slp$(f,\mtt{unmark})$\}}
\LeftLabel{[cons]}
\UnaryInfC{\{$f$\}$~\mtt{unmark}~$\{$f$\}}
\LeftLabel{[alap]}
\UnaryInfC{\{$f$\}$~\mtt{unmark!}~$\{$f\wedge\text{Fail}(\mtt{unmark})$\}}
\LeftLabel{[cons]}
\UnaryInfC{\{$e\wedge\text{Success}(\mtt{Illegal})$\}$~\mtt{unmark!}~$\{$c\vee d$\}}

\AxiomC{}
\LeftLabel{[ruleapp]$_\text{slp}$}
\UnaryInfC{\{$d$\}$~\mtt{skip}~$\{$d$\})}
\LeftLabel{[cons]}
\UnaryInfC{\{$e\wedge\text{Fail}(\mtt{Illegal})$\}$~\mtt{skip}~$\{$c\vee d$\}}

\LeftLabel{[if]}
\BinaryInfC{\{$e$\}$~\mtt{if~Illegal~then~umark!}~$\{$c\vee d$\}}
\end{prooftree}
\end{scriptsize}
    \caption{Proof tree for partial correctness of $\mtt{2colouring}$}
    \label{fig:2col-proof}
\end{figure}

Note that there is no command $\mtt{break}$ in the program, so Break$(c,P,\mrm{false})$ always holds regardless $c$ and $P$ for this program. For this reason and for simplicity, we omit premise Break$(c,P,\mrm{false})$ in the inference rule [alap] of the proof tree.

For an example of constructing Slp, let us consider the rule $r=\mtt{init}$ of program $\mtt{2-colouring}$ and the formula $f$ of Table 2. Note that $\forall x(c)$ is an abbreviation of $\neg\E{}x(\neg c)$ so that we need to change universal quantifiers to existential quantifiers.\\
\begin{footnotesize}\begin{tabular}{lcl}\footnotesize
{Split}$(f,r)$&
=&$\mrm{\neg((\mV(1)\neq red\wedge\mV(1)\neq blue\wedge\mV(1)\neq none)\vee root(1))}$\\
&&$\mrm{\wedge\neg\E{V}x(x\neq 1\wedge(\mV(x)\neq red\wedge\mV(x)\neq blue\wedge\mV(x)\neq none)\vee root(x))}$\\
&&$\mrm{\wedge\neg\E{E}x(\mE(x)\neq none)}$
\end{tabular}\\
\noindent Dang$(r)$ = $\mrm{true}$\\
\begin{tabular}{lcl}\footnotesize
{Lift}$(f,r^\vee)$&=&
$\mrm{\neg\E{V}x(x\neq 1\wedge(\mV(x)\neq red\wedge\mV(x)\neq blue\wedge\mV(x)\neq none)\vee root(x))}$\\
&&$\mrm{\wedge\neg\E{E}x(\mE(x)\neq none)}$
\end{tabular}\\
{Adj}(Lift$(f,r^\vee),r)$ = Lift$(f,r^\vee)$\\
Shift$(f,r^\vee)$=Lift$(f,r^\vee)\,\wedge\,\mrm{\lV(1)=a\wedge\mV(1)=red\wedge\neg root(1)}$\\
\begin{tabular}{lcl}
Slp$(f,r)$&$\equiv$&$\mrm{\E{V}y(\neg\E{V}x(x\neq y\wedge(\mV(x)\neq red\wedge\mV(x)\neq blue\wedge\mV(x)\neq none)\vee root(x))}$\\
&&$\mrm{\neg\E{E}x(\mE(x)=none)\wedge\E{L}a(\lV(y)=a)\wedge\mV(y)=red\wedge\neg root(y))}$
\end{tabular}
\end{footnotesize}

In the proof tree of \figurename~\ref{fig:2col-proof}, we apply some inference rule [cons] which means we need to give proof of implications applied to the rules. Some implications are obvious, e.g. $c$ implies $c\vee d$. Other implications, are also obvious if we check their formulas. The implications have the form $\exists y(\forall x((x=y\vee c)\wedge x=y\Rightarrow c))$ for some variables $x,y$ and FO formula $c$ with no variable $y$, which implies $\forall x(c)$. For an example, Post$(f,\mtt{init})$ expresses that there exists an unrooted red node $y$, labelled with a list, where all nodes beside $y$ are unmarked or marked red or blue, which implies all nodes are unmarked or marked red or blue, such that $f$ holds. Other proof of implications use a similar method (see \cite{WP20a}).

\section{Soundness and Completeness of the Proof Calculi}
\label{sec:completeness}
In \cite{WP20a}, we show that both \textsf{SEM} and \textsf{SYN} are sound. That is, if a triple $\{c\}~P~\{d\}$ can be proven by \textsf{SEM} or \textsf{SYN} (denoted by $\vdash_{\textsf{SEM}}$ or $\vdash_{\textsf{SYN}}$), then the triple is partially correct.

\begin{theorem}[Soundness]\label{theo:semsound}\normalfont
Given a graph program $P$ and assertions $c,d$. Then, $\vdash_{\mathsf{SEM}}\{c\}~P~\{d\} \text{~implies~} \vDash \{c\}~P~\{d\}$. Moreover, if $c$ and $d$ are first-order formulas, $\vdash_{\mathsf{SYN}}\{c\}~P~\{d\} \text{~implies~} \vDash \{c\}~P~\{d\}$.
\end{theorem}

A proof calculus is complete if every partially correct triple can be proved by the calculus. 
Neither \textsf{SEM} nor \textsf{SYN} are complete because GP\,2's expressions include Peano arithmetic which is known to be incomplete \cite{Monk76a}.
However, the notion of relative completeness allows to separate the incompleteness in proving valid assertions from the power of the inference rules for programming constructs \cite{Cook78}. That means, we assume that the implications in the [cons] rules of SEM and SYN can be proved outside the calculi.

\Theo{Relative completeness of \textsf{SEM}}{theo:semcomplete}{
Given a graph program $P$ and assertions $c,d$. Then, $\vDash\{c\}~P~\{d\} \text{~implies~} \vdash_{\mathsf{SEM}} \{c\}~P~\{d\}$.}

The proof of Theorem \ref{theo:semcomplete} can be seen in \cite{WP20a}. The proof relies on the existence of WLP($P,c$) for arbitrary programs $P$ and assertions $c$. Even if we omit [ruleapp$_{\text{slp}}$] from the calculus, \textsf{SEM} is still relative complete. However, for \textsf{SYN} to be relative complete, it would be necessary to express WLP($P,c$) or SLP ($c,P$) as first-order formulas. There is strong evidence that this is impossible. For example, consider the triple $\{c\}~P~\{d\}$ with $c=\mrm{\A{V}x(m_V(x)=none\land\neg\E{E}y(s(y)=x\vee t(y)=x))}$ 
(all nodes are unmarked and isolated), $d=\mrm{\A{V}x(false)}$ (the graph is empty), and the following program:

\vspace{1ex}
\begin{samepage}
\begin{small}
$~~~~~~~~~~~~~~~~~~~~~~~~~~~~~~~~~~~~~~~~~~\mtt{Main = duplicate!;\, delete!}$\\[1ex]
$~~~~~~~~~~~~~~~~~~~~~~~~~~~~~~~~~~~~~~~~~~~~~~~~$\begin{tabular}[t]{lll}
\begin{tikzpicture}[remember picture,
  inner/.style={circle,draw,minimum size=16pt},
  outer/.style={inner sep=2pt}, scale=0.7
  ]
  \node[outer] (AA) at (0,1) {$\tiny\mtt{duplicate(a:list)}$};
  \node[outer] (A) at (-1,0) {
   \begin{tikzpicture}[scale=0.7, transform shape]
		\node[inner, label=below:\tiny 1] (Aa) at (0,0) {$\tiny\mtt{a}$};
		\end{tikzpicture}};
  \node[outer] (B) at (0,0) {$\Rightarrow$};
  \node[outer] (C) at (1.5,0) {
   \begin{tikzpicture}[scale=0.7, transform shape]
		\node[inner, label=below:\tiny 1, fill=gray!50] (Aa) at (0,0) {$\tiny\mtt{a}$};	
		\node[inner, fill=gray!50] (Aa) at (1,0) {$\tiny\mtt{a}$};	
		\end{tikzpicture}};
\end{tikzpicture}
&\hspace{5cm}&
\begin{tikzpicture}[remember picture,
  inner/.style={circle,draw,minimum size=16pt},
  outer/.style={inner sep=2pt}, scale=0.7
  ]
  \node[outer] (AA) at (-1,1) {$\tiny\mtt{delete(a:list)}$};
  \node[outer] (A) at (-1.5,0) {
   \begin{tikzpicture}[scale=0.7, transform shape]
		\node[inner, fill=gray!50] (Aa) at (0,0) {$\tiny\mtt{a}$};	
		\node[inner, fill=gray!50] (Aa) at (1,0) {$\tiny\mtt{a}$};	
		\end{tikzpicture}};
  \node[outer] (B) at (0,0) {$\Rightarrow$};
  \node[outer] (C) at (1,0) {$\emptyset$};
\end{tikzpicture}
\end{tabular}
\end{small}
\end{samepage}

It is obvious that $\vDash \{c\}~\mtt{duplicate!; delete!}~\{d\}$ holds: $\mtt{duplicate!}$ duplicates the number of nodes while marking the nodes grey, hence its result graph consists of an even number of isolated grey nodes. Then $\mtt{delete!}$ deletes pairs of grey nodes as long as possible, so the overall result is the empty graph. Note that ``consists of an even number of isolated grey nodes" is both the strongest postcondition with respect to $c$ and \ttt{duplicate!}, and the weakest precondition with respect to \ttt{delete!} and $d$.

Using \textsf{SYN} one can prove $\vdash \{c\}~\mtt{duplicate!}~\{e\}$ where $e$ expresses that all nodes are grey and isolated. However, we believe that our logic cannot express that a graph has an even number of nodes. This is because pure first-order logic (without built-in operations) cannot express this property \cite{Libkin04} and it is likely that this inexpressiveness carries over to our logic. As a consequence, one can only prove $\vdash \{e\}~\mtt{delete!}~\{f\}$ where $f$ expresses that the graph contains at most one node (because otherwise \ttt{delete} would be applicable). But we cannot use \textsf{SYN} to prove  $\vdash \{c\}~\mtt{duplicate!; delete!}~\{d\}$.

\section{Related Work}
\label{sec:related_work}
Hoare-style verification of graph programs with attributed rules was introduced in \cite{PoskittP12,Poskitt13}, using E-conditions which generalise the nested graph conditions of Habel and Pennemann \cite{HP09,Pennemann09}. E-conditions do not cover rooted rules or the $\mtt{break}$ command, which are considered in our first-order formulas. More importantly, the approach of \cite{PoskittP12,Poskitt13} can only handle programs in which the conditions of branching commands and loop bodies are rule set calls. Our syntactic calculus \textsf{SYN} covers a larger class of graph programs, viz.\ programs where the condition of each branching command is a loop-free program, and each loop body is an iteration command. This allows us, in particular, to verify many programs with nested loops. Besides this increased power, we believe that assertions in the form of first-order formulas are easier to comprehend by programmers than nested graph conditions of some form. 


As argued at the end of the previous section, we cannot express SLP$(c,P)$ or WLP$(P,c)$ for arbitrary assertions $c$ and graph programs $P$ as first-order formulas. In \cite{HP09,Pennemann09}, there is a construction of Wlp$(c,P!)$ by using an infinite formula. Here, we do not use a similar trick but stick to standard finitary logic. The papers \cite{DijkstraS90,Jones-Roscoe-Wood10a} do not give constructions for syntactic strongest liberal postconditions or weakest liberal postconditions either. Instead, similar to the consequent of our inference rule [alap], the conjunction of a loop invariant and a negated loop condition is considered as an ``approximate" strongest liberal postcondition. 

In \cite{Brenas-Echahed-Strecker18b}, the authors design an imperative programming language for manipulating graphs and give a Hoare calculus based on weakest preconditions. Programs manipulate the graph structure only and do not contain arithmetic. Assertions are formulas of the so-called guarded fragment of first-order logic, which is decidable. This relatively weak logic makes the correctness of programs decidable. 

Our goal is different in that we want a powerful assertion language that can specify many practical algorithms on graphs. (In fact, we plan to extend our logic to monadic second-order logic in order to express non-local properties such as connectedness, colourability, etc.) In our setting, it is easily seen that correctness is undecidable in general, even for trivial programs. For example, consider Hoare triples of the form $\{\mrm{true}\} \mtt{skip} \{d\}$ where d is an arithmetic formula (without references to nodes or edges). Such a triple is partially (and totally) correct if and only if d is true on the integers. But our formulas include Peano arithmetic and hence are undecidable in general \cite{Monk76a}. Thus, even for triples of the restricted form above, correctness is undecidable.

\section{Conclusion and Future Work}
\label{sec:conclusion}
We have shown how to construct a strongest liberal postcondition for a given conditional rule schema and a precondition in the form of a first-order formula. Using this construction, we have shown that we can obtain a strongest liberal postcondition over a loop-free program, and construct a first-order formula for SUCCESS$(C)$ for a loop-free program $C$. Moreover, we can construct a first-order formula for FAIL$(P)$ for an iteration command $P$. Altogether, this gives us a proof calculus that can handle more programs than previous calculi in the literature, in particular we can now handle certain nested loops.

However, the expressiveness of first-order formulas over the domain of graphs is quite limited. For example, one cannot specify that a graph is connected by a first-order formula. Hence, in the near future, we will extend our formulas to monadic second-order formulas to overcome such limitations \cite{Cou12}. 

Another limitation in current approaches to graph program verification is the inability to specify isomorphisms between the initial and final graphs \cite{WP18}. Monadic second-order transductions can link initial and final states by expressing the final state through elements of the initial state \cite{Cou12}. We plan to adopt this technique for graph program verification in the future. 

\bibliographystyle{eptcs}
\bibliography{firstorder}

\end{document}